\documentclass[amsmath,amssymb,prb,aps,showpacs,showkeys,twocolumn]{revtex4}
\usepackage{bm,graphicx,subfigure,enumerate,latexsym,comment}
\usepackage[dvips]{epsfig}
\usepackage{amsmath,amssymb}

\newcommand{\bol}[1]{\boldsymbol{\mathrm{#1}}}
\newcommand{\derivative}[2]{\frac{{\rm d} #1}{{\rm d} #2}}
\newcommand{\del}{\partial}
\renewcommand{\tilde}{\widetilde}

\begin{document}

\title{Multiscale Simulations for Polymeric Flow}
\author{Takahiro Murashima
\footnote{Electronic mail: murasima@cheme.kyoto-u.ac.jp}}
\author{Takashi Taniguchi
\footnote{Electronic mail: taniguch@cheme.kyoto-u.ac.jp}}
\author{Ryoichi Yamamoto
\footnote{Electronic mail: ryoichi@cheme.kyoto-u.ac.jp}}
\author{Shugo Yasuda
\footnote{Electronic mail: yasuda@cheme.kyoto-u.ac.jp}}

\affiliation{
Department of Chemical Engineering, 
Kyoto University, Kyoto 615-8510, Japan, and\\
CREST, Japan Science and Technology Agency, Kawaguchi 332-0012, Japan.
}
\date{\today}

\begin{abstract}
Multiscale simulation methods have been developed based on the local stress sampling strategy and applied to three flow problems with different difficulty levels: (a) general flow problems of simple fluids, (b) parallel (one-dimensional) flow problems of polymeric liquids, and (c) general (two- or three-dimensional) flow problems of polymeric liquids. In our multiscale methods, the local stress of each fluid element is calculated directly by performing microscopic or mesoscopic simulations according to the local flow quantities instead of using any constitutive relations. For simple fluids (a), such as the Lenard-Jones liquid, a multiscale method combining MD and CFD simulations is developed based on the local equilibrium assumption without memories of the flow history. The results of the multiscale simulations are compared with the corresponding results of CFD with or without thermal fluctuations. The detailed properties of fluctuations arising in the multiscale simulations are also investigated. For polymeric liquids in parallel flows (b), the multiscale method is extended to take into account the memory effects that arise in hydrodynamic stress due to the slow relaxation of polymer-chain conformations. The memory of polymer dynamics on each fluid element is thus resolved by performing MD simulations in which cells are fixed at the mesh nodes of the CFD simulations. The complicated viscoelastic flow behaviours of a polymeric liquid confined between oscillating plates are simulated using the multiscale method. For general (two- or three-dimensional) flow problems of polymeric liquids (c), it is necessary to trace the history of microscopic information such as polymer-chain conformation, which carries the memories of past flow history, along the streamline of each fluid element. A Lagrangian-based CFD is thus implemented to correctly advect the polymer-chain conformation consistently with the flow. On each fluid element, coarse-grained polymer simulations are carried out to consider the dynamics of entangled polymer chains that show extremely slow relaxation compared to microscopic time scales. This method is successfully applied to simulate a flow passing through a cylindrical obstacle.
\end{abstract}



\maketitle

\section{introduction}
The prediction of flow behaviours of polymeric liquids using computer simulations is a challenging theme in various fields of science and engineering including physical science, materials science, and mechanical and chemical engineering. 
Polymeric liquids are known to exhibit peculiar flow behaviours that are related to the microscale dynamics of their polymer chains; these dynamics affect the viscoelasticity, shear thinning/thickening behaviours, and flow-induced phase transitions of these liquids.\cite{book:87BAH}
The characteristic times of the microscale dynamics of polymer chains tend to be very long and are often comparable to the time scales of macroscale fluid motions. 
Thus, for these compounds, one cannot separate the microscale and macroscale dynamics in the temporal domain, even though a large gap exists between those length and time scales; one also cannot make the assumption usually made for simple fluids under flow that the fluid elements are in local equilibrium at each instance. 
Therefore, one must trace the ``memory'' or ``history'' of the deformations of each fluid element along its streamline. 
This coupling between microscale and macroscale dynamics hinders the simulation of polymeric liquids. 

In this paper, we present a detailed explanation of a multiscale method that bridges the hydrodynamic motions of fluids using computational fluid dynamics (CFD) and the microscopic [or mesoscopic] dynamics of polymer configurations using molecular-dynamics (MD) [or coarse-grained (CG)\cite{SLTM2001,TTD2001,MTKIMG2001,DT2003,art:09IMO}] simulations. 
The concept of bridging microscale and macroscale dynamics is also important for other flow problems of softmatters with complex internal degrees of freedom (e.g., colloidal dispersion, liquid crystal, and glass). 
The basic idea of our multiscale method can be applied to those softmatter flows.

\begin{figure*}[t]
\begin{center}
\includegraphics[scale=0.65]{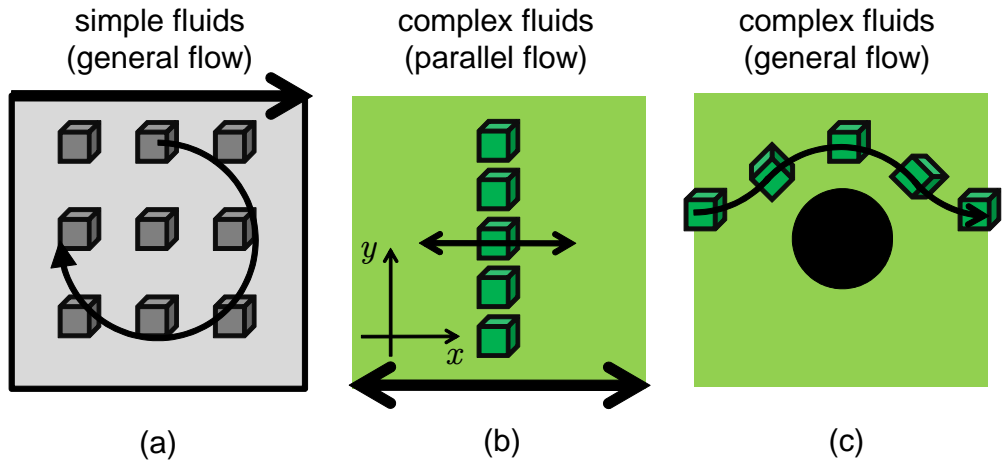}
\end{center}
\caption{
Schematic illustrations of our multiscale methods bridging CFD and MD [or CG] simulations. The developments were carried out in a step-by-step manner starting from the bridging method for simple fluids (a) and proceeding to a method applicable to general flow problems of complex fluids. The small cubic boxes placed in the fluid systems represent MD or CG simulation cells within which the local stress of each fluid element is sampled. Case (a) represents a cavity flow of a simple fluid for which the local stress $\sigma$ of the fluid at a time $t$ and a position ${\bm r}$ is given by a function of the local deformation rate $\dot{\gamma}$ at the same time and position, $\sigma({\bm r},t)=f(\dot{\gamma}({\bm r},t))$. Case (b) represents a polymeric liquid subject to an oscillatory shear flow where the flow is parallel to the $x$-direction and the velocity gradient exists only in the $y$-direction. Here, the local stress is a function of the history of the velocity gradient in the past $t'\leq t$ at the position $y$, $\sigma(y,t)=F[\dot{\gamma}(y,t');~t'\leq t]$. Case (c) represents a polymeric liquid passing through a cylindrical obstacle. Here, the local stress is a function of the history of the velocity gradient in the past $t'\leq t$ along the streamline ${\bm R}(t')$ of the fluid element, $\sigma({\bm r},t)=F[\dot{\gamma}({\bm R}(t'),t');~t'\leq t]$. The details of our multiscale methods for cases (a), (b), and (c) are given in Secs. II, III, and IV, respectively.
}
\label{mss}
\end{figure*}

In non-Newtonian fluid dynamics, many novel CFD schemes have been proposed for the viscoelastic flows of polymeric liquids.\cite{art:86LTI,art:86LTII,art:90RAB,art:95GF,art:98B} 
In CFD methods, a model constitutive equation is used to determine the local stresses at each instant from the history of previous velocity fields.
For dense polymeric liquids (melts), however, the detailed form of the constitutive relations is so complicated that they are generally unknown.\cite{book:88L}
Thus, the usual CFD methods, which require constitutive equations, are usually not straightforwardly applicable to the complicated flow problems of polymeric liquids. 
Instead, microscopic simulations such as MD and CG simulations are often used to investigate the rheological properties of such materials. The microscopic simulations are usually performed only for a tiny piece of the material in the equilibrium or non-equilibrium state under uniform external fields of shear velocity, temperature gradient, and electric field. 
Although the microscopic simulations are even applicable to macroscopic flow behaviours, the drawback of this type of simulation is the enormous computation time required. 
Thus, for problems that concern large-scale and long-time fluid motions far beyond the molecular scale, which are commonly encountered in engineering, fully microscopic simulations are difficult from a practical standpoint. 
To overcome the weaknesses of the individual methods, we have developed a multiscale simulation method that combines CFD and MD [or CG] simulations.\cite{art:08YY,art:09YY,art:10YY,MT2010a,MT2010b}

In our multiscale methods, macroscopic flow behaviour is calculated by a CFD solver; however, instead of using any constitutive equations, a local stress is calculated using the MD [or CG] simulation associated with a local fluid element according to the local flow variable obtained by the CFD calculation. 
The multiscale methods are applied to three flow problems with different levels of difficulty: (a) general flow problems of simple fluids, (b) parallel (one-dimensional) flow problems of polymeric liquids, and (c) general (two- or three-dimensional) flow problems of polymeric liquids, as schematically illustrated in Fig.~\ref{mss}.

For the simple fluids shown in Fig. ~\ref{mss} (a), the local stress $\sigma$ of the fluid at a time $t$ and a position ${\bm r}$ is given by a function of the local deformation rate $\dot{\gamma}$ at the same time and position,
\begin{equation}
\sigma({\bm r},t)=f(\dot{\gamma}({\bm r},t)).
\end{equation}
We proposed a multiscale method that combines CFD and MD based on the local equilibrium assumption, in which a local stress immediately attains a steady state after a short transient time during which a strain rate is subjected to the fluid element. 
In this method, a lattice-mesh-based CFD scheme is used at the macroscopic level, and the non-equilibrium MD simulations are performed in small MD cells associated with each lattice node of the CFD to generate a local stress according to a local strain rate.

The multiscale method combining CFD and MD is extended in a straightforward fashion to the polymeric flows in the one-dimensional geometry shown in Fig.~\ref{mss} (b), in which the macroscopic quantities, e.g., velocity, temperature, and stress, are uniform in the flow direction parallel to the plates. 
This situation allows us to neglect the advection of memory on a fluid element along the streamline, so the local stress of a fluid element at $t$ and $y$ is a functional of the history of the velocity gradient in the past $t'\leq t$ at the same position $y$,
\begin{equation}
\sigma(y,t)=F[\dot{\gamma}(y,t');~t'\leq t].
\end{equation}

For general two- or three-dimensional flow problems of polymeric liquids, shown in Fig.~\ref{mss} (c), one must consider the advection of polymer-chain conformations, which carries memory effects on a fluid element along the streamline. 
The local stress is thus a functional of the history of the velocity gradient in the past $t'\leq t$ along the streamline ${\bm R}(t')$ of the fluid element,
\begin{equation}
\sigma({\bm r},t)=F[\dot{\gamma}({\bm R}(t'),t');~t'\leq t].
\end{equation}
To meet this requirement, Lagrangian fluid dynamics are implemented for the CFD calculation to trace the advection of a fluid element that contains the memory of the configuration of the polymer chains. 
The local stresses on each fluid particle are calculated using coarse-grained polymer simulations in which the dynamics of entangled polymer chains are calculated.

The idea of using multiscale modelling to calculate the local stress for the fluid solver using the microscopic simulation instead of any constitutive relations was first proposed for polymeric liquids by Laso and \"Ottinger\cite{art:93LO, art:95FLO, art:97LPO}, who presented the CONNFFESSIT approach. 
The multiscale method was also proposed by E and Enquist,\cite{art:03EE} who presented the heterogeneous multiscale method (HMM) as a general methodology for the efficient numerical computation of problems with multiscale characteristics. 
HMM has been applied to the simulation of complex fluids\cite{art:05RE,art:07EELRV} but completely neglects the advection of memory. 
The equation-free multiscale computation proposed by Kevrekidis {\it et al}. is based on a similar idea and has been applied to various problems.\cite{art:03KGHKRT,art:09KS}
The basic idea of our multiscale modelling is the same as those earlier proposed. 
This type of bridging method has also been developed recently by several researchers. 
De {\it et al.} have proposed the scale-bridging method, which can correctly reproduce the memory effect of a polymeric liquid, and demonstrated the non-linear viscoelastic behaviour of a polymeric liquid between oscillating plates.\cite{art:06DFSKK}
The methodology of the present multiscale simulations for one-dimensional flows of polymeric liquids is the same as that used in the scale-bridging method. 
Kessler {\it et al.} have recently developed a multiscale simulation for rarefied gas flows based on a similar idea.\cite{art:10KOK}

In the following, the bridging method of MD and CFD simulations for simple fluids is presented in Sec. II. 
In this section, the results of the 2-D cavity flows are compared with those of Newtonian fluids to demonstrate the validity of our multiscale simulation method. 
A special focus is placed on the fluctuation arising in our multiscale method. 
We carry out spectral analysis of the fluctuations and compare the results with those obtained using fluctuating hydrodynamics. 
In Sec. III, the multiscale method of MD and CFD simulations is extended to the one-dimensional flows of polymeric liquids confined in parallel plates. 
The flow behaviours of a glassy polymer melt in the oscillating plates are calculated, the local rheological properties of the polymer melt in the rapidly oscillating plates are investigated, and the results are compared with the analytical solution for an infinitesimally small strain to demonstrate the validity of the multiscale method. 
In Sec. IV, the multiscale method of Lagrangian dynamics and CG polymer-dynamics simulations is presented for a two-dimensional flow problem of polymeric liquids. 
A transient flow passing a circular object is demonstrated. Summaries of each section and future perspectives are given in Sec. V.

\section{Simple Fluid}
We consider a simple liquid composed of particles interacting via the repulsive part of the Lennard-Jones (LJ) potential,
\begin{equation}\label{sec2_eq0}
U_{\rm LJ}(r)=
\left\{
\begin{array}{c c}
4\epsilon\left[
({\sigma}/{r})^{12}
-({\sigma}/{r})^{6}
\right]
+\epsilon & (r\le 2^{1/6}\sigma),\\
0 & ( r> 2^{1/6}\sigma).
\end{array}
\right. ,
\end{equation}
where $\sigma$ and $\epsilon$ are the length and energy units of the LJ potential, respectively. 
In this section, we measure space and time in the units of $\sigma$ and $\tau_0=\sqrt{m\sigma^2/\epsilon}$, respectively. 
The temperature $T$ is measured in the unit of $\epsilon/k_B$. 
Here, $m$ and $k_B$ are the mass of the LJ particle and the Boltzmann constant, respectively. 
The temperature $T$ and density $\rho$ of the liquid are assumed to be uniform and fixed as $T=1.0$ and $\rho=0.8$.

\begin{figure}[t]
\includegraphics[scale=1]{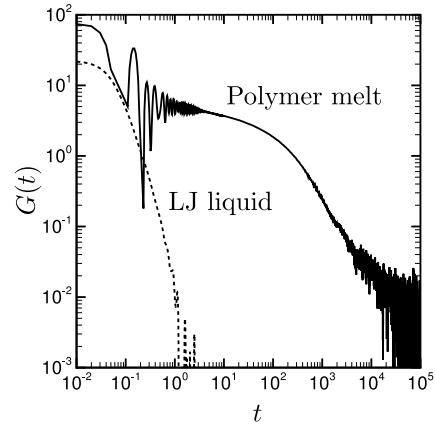}
\caption{
The stress relaxation function $G(t)$ for the LJ liquid (dashed line) and model polymer melt (solid line).
}\label{sec2_fig_gt}
\end{figure}

At this density and temperature, the LJ liquid does not have a long-time memory. 
That is, the shear stress depends only on the instantaneous strain rate, regardless of the history of the previous strain rates experienced by the fluid element. 
In Fig. \ref{sec2_fig_gt}, we show the stress-relaxation function $G(t)$ of the present LJ liquid and of a model polymeric liquid (the latter will be discussed in the next section). 
The stress-relaxation function $G(t)$ is calculated as
\begin{equation}
G(t)=\langle\, \Pi_{xy}(t+t_0)\Pi(t_0)\,\rangle /k_B T V,
\end{equation}
where $\Pi_{xy}$ is the space integral of the microscopic stress tensor in the volume $V$. 
It is seen that the stress relaxation of the LJ liquid rapidly decreases and that the time correlation of stress almost disappears in $t \ll 1$. 
The relaxation time $\tau_{LJ}$ of the LJ liquid may be estimated as $\tau_{LJ}=0.067$ with the definition $G(\tau_{LJ})/G(0)=e^{-1}$. 
The time scale of temporal variations in the macroscopic flows may be much larger than the stress-relaxation time of the present LJ liquid. 
Thus, for the macroscopic flow behaviours of simple fluids, one can ignore the memory effect in a local stress due to the history of previous flow velocities. 
That is, we can assume the local equilibrium states at any instant in the macroscopic flows. 

In this section, we present a model for multiscale simulations of MD and CFD for simple fluids without memory effects. 
The bridging scheme of MD and CFD is based on the local equilibrium assumption of the macroscopic quantities. 
The multiscale simulations are performed for the driven cavity flows of the simple LJ liquid, and the results are compared with those of the Newtonian fluid. 
Special attention is given to the efficiency of the present multiscale method and to the noise arising in this method.

\subsection{Multiscale Model}
Incompressible flows with uniform density $\rho_0$ and temperature $T_0$ are described by the following equations:
\begin{equation}\label{sec2_eq1}
\frac{\partial v_\alpha}{\partial x_\alpha} = 0,
\end{equation}
\begin{equation}\label{sec2_eq2}
\frac{\partial v_\alpha}{\partial t} + v_\beta\frac{\partial v_\alpha}{\partial x_\beta}
=\frac{1}{\rho_0}\frac{\partial P_{\alpha\beta}}{\partial x_\beta} + g_\alpha,
\end{equation}
where $x_\alpha$ is the Cartesian coordinate system, $t$ the time, $v_\alpha$ the velocity, $\rho_0$ the density, $P_{\alpha\beta}$ the stress tensor, and $g_\alpha$ the external force per unit mass. 
Throughout this work, the subscripts $\alpha$, $\beta$, and $\gamma$ represent the index in Cartesian coordinates, i.e., \{$\alpha$, $\beta$, $\gamma$\} = \{$x$, $y$, $z$\}, and the summation convention is used. 
The stress tensor $P_{\alpha\beta}$ is written in the form
\begin{equation}\label{sec2_eq3}
P_{\alpha\beta} = -p \delta_{\alpha\beta} + T_{\alpha\beta},
\end{equation}
where $p$ is the pressure and $\delta_{\alpha\beta}$ is the Kronecker delta. 
We may assume that $T_{\alpha\beta}$ is symmetric, $T_{\alpha\beta}=T_{\beta\alpha}$, and traceless, $T_{\alpha\alpha}$=0, for isotropic simple fluids.\cite{art:45R}
To solve the above equations, one needs a constitutive relation for the stress tensor $T_{\alpha\beta}$. 
In our multiscale method, instead of using any explicit formulas such as the Newtonian constitutive relation, $T_{\alpha\beta}$ is computed directly by MD simulations.

\subsubsection{CFD Scheme}
\begin{figure*}[htbp]
\begin{center}
\includegraphics[scale=1]{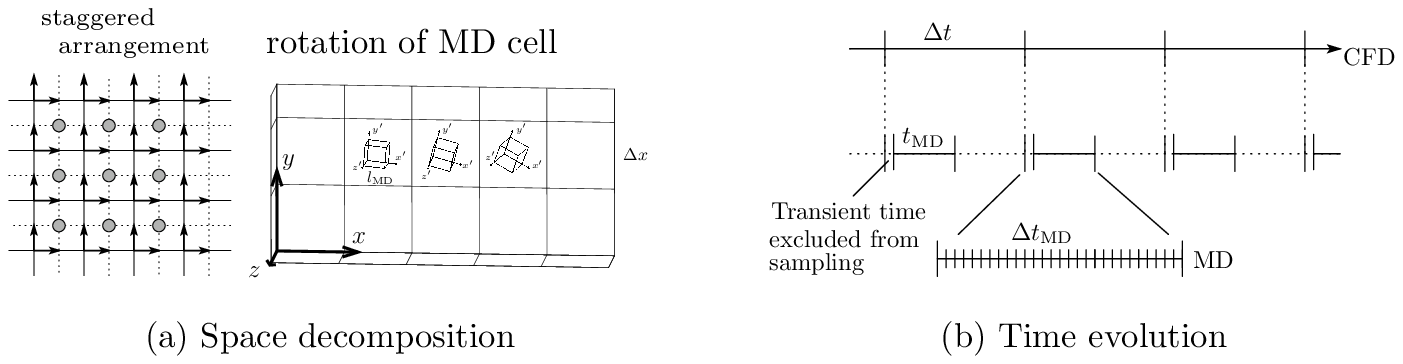}
\end{center}
\caption{
Schematic diagram of the multiscale simulation method for simple fluids. (a) Staggered arrangement of the velocity $\bm v$ and stress tensor $P_{\alpha\beta}$ on a lattice-mesh grid of the CFD calculation. CFD simulations are performed in a reference coordinate ($x$,$y$,$z$), while MD simulations are performed in a rotated coordinate ($x'$,$y'$,$z'$) so that the diagonal components of $E'_{\alpha\beta}$ all become zero using the procedure described in Sec II B. The CFD system is discretised into cubic subsystems whose side length is $\Delta x$. Each subsystem is associated with an MD cell whose side length is $l_{\rm MD}$, with the Lees-Edward periodic boundary condition under shear deformation. Note that three-dimensional MD simulations are used at the microscopic level even for the problems for which one- or two-dimensional analysis is applied at the macroscopic level. (b) A schematic time evolution of our multi-scale method. The CFD simulation proceeds with a time step of $\Delta t$, and the MD simulation is carried out for a lapse of time $t_{\rm MD}$ only to sample the local stress $T'_{\alpha\beta}$ at each node point and time step of CFD.
}
\label{sec2_f00}
\end{figure*}

We use a lattice-mesh-based finite-volume method with a staggered arrangement for vector and scalar quantities\cite{book:02FP} (Fig.~\ref{sec2_f00}). 
The control volume for a vector quantity is a unit square surrounded by dashed lines, and that for a scalar quantity is a unit square surrounded by solid lines. 
Equations. (\ref{sec2_eq1}) and (\ref{sec2_eq2}) are discretised by integrating the quantities on each control volume. 
For numerical time integrations, we use the fourth-order Runge-Kutta method, in which a single physical time step $\Delta t$ is divided into four sub-steps. 
The time evolution of a quantity $\phi$, which is to be determined by the equation $\partial \phi/\partial t$=$f(t,\phi)$, is written as
\begin{subequations}\label{sec2_eq4}
\begin{align}
\phi_{n+\frac{1}{2}}^*&=\phi^n + \frac{\Delta t}{2}f(t_n,\phi^n),
\\
\phi_{n+\frac{1}{2}}^{**}&=\phi^n + \frac{\Delta t}{2}f(t_{n+\frac{1}{2}},\phi^*_{n+\frac{1}{2}}),
\\
\phi_{n+1}^{*}&=\phi^n + \Delta t f(t_{n+\frac{1}{2}},\phi^{**}_{n+\frac{1}{2}}),
\\
\phi^{n+1}&=\phi^n + \frac{\Delta t}{6}\left[
f(t_{n},\phi^{n}) + 2 f(t_{n+\frac{1}{2}},\phi^*_{n+\frac{1}{2}}) +
\right.
\nonumber
\\
&\hspace{2cm} 
\left. 2 f(t_{n+1/2},\phi^{**}_{n+1/2}) + f(t_{n+1},\phi^*_{n+1}) 
\right].
\end{align}
\end{subequations}
The time evolution of the fluid velocity $\bm v$ is computed by the above set of equations. 
On the other hand, the pressure $p$ is determined so that the fluid velocity satisfies the incompressible condition (\ref{sec2_eq1}) at each sub-step. 
The procedure at each sub-step is written as
\begin{subequations}\label{sec_eq5}
\begin{align}
&p = \tilde p + \psi,
\\
&\bm v =\tilde {\bm v} -\tau \nabla \psi,
\\
&\triangle \psi = \frac{1}{\tau}\nabla{\tilde {\bm v}},
\end{align}
\end{subequations}
where $\tilde p$ is the pressure obtained at the previous sub-step, $\tilde {\bm v}$ is the velocity obtained by solving equation (\ref{sec2_eq4}) at the present sub-step, $\tau$ is the time increment of the sub-step, and $\psi$ is the correction term to obtain the divergence-free velocities. 
The remaining three components of the stress tensor, $T_{\alpha\beta}$, are to be computed directly by MD simulations. 
The details of the method are described in the next subsection. 
Note that the calculation of $T_{\alpha\beta}$ is carried out at each sub-step of Eq. (\ref{sec2_eq4}).

\subsubsection{Local MD sampling}
We compute the local stresses by MD simulations according to the local strain rates, rather than the local flow velocities themselves. A schematic diagram of the method is depicted in Fig. \ref{sec2_f00}. 
At the CFD level, the local strain rate tensor $E_{\alpha\beta}$ is defined as
\begin{equation}\label{sec2_eq6}
E_{\alpha\beta}=\frac{1}{2}\left (\frac{\partial v_\alpha}{\partial x_\beta} + \frac{\partial v_\beta}{\partial x_\alpha}\right ),
\end{equation}
where the incompressible condition, $E_{\alpha\alpha}$=0, is to be satisfied. 
We can now define a rotation matrix $\Theta$ by which the strain-rate tensor $E_{\alpha\beta}$ is transformed to
\begin{equation}\label{sec2_eq7}
E'=\Theta E \Theta ^{\rm T} =\left(
\begin{array}{ccc}
0&E'_{xy}&E'_{xz}\\
E'_{yx}&0&E'_{yz}\\
E'_{zx}&E'_{zy}&0
\end{array}
\right),
\end{equation}
where the diagonal components all vanish. 
This transformation enables us to perform MD simulations with the Lees-Edwards periodic boundary condition.

We note that the Lees-Edwards periodic boundary condition cannot create a flow field for an arbitrary velocity-gradient tensor $\partial v_\alpha /\partial x_\beta$ in an MD cell but can reproduce a velocity profile for an arbitrary (symmetric) strain-rate tensor $E_{\alpha\beta}$ using three components of the velocity-gradient tensor, e.g., $\partial v_x/\partial y$, $\partial v_z/\partial y$, and $\partial v_x/\partial z$. 
For simple fluids, the antisymmetric part of the velocity-gradient tensor, $\Omega_{\alpha\beta}=\frac{1}{2}(\partial v_\alpha /\partial x_\beta - \partial v_\beta /x_\alpha)$, does not affect the stress tensor $T_{\alpha\beta}$. 
Thus, the present multiscale method using the Lees-Edwards boundary condition in each MD cell is applicable to the general (three-dimensional) flows of simple fluids.

The off-diagonal stress tensor $T'_{\alpha\beta}$ is computed according to $E'_{\alpha\beta}$ and then passed to CFD after transforming back into the original coordinates, $T_{\alpha\beta}$. 
For two-dimensional flows [$\partial/\partial z $=0 and $v_z$=0], $\Theta$ and $E'$ are expressed as
\begin{align}
\Theta&=\left(
\begin{array}{cc}
 \cos\theta& \sin\theta\\
-\sin\theta& \cos\theta
\end{array}
\right),
\label{sec2_eq8}
\\
E'_{xy}=E'_{yx}&=-E_{xx}\sin 2\theta + E_{xy}
\cos 2\theta,
\label{sec2_eq9}
\end{align}
where
\begin{equation}\label{sec2_eq10}
\theta=\frac{1}{2}\tan^{-1}
\left(-\frac{E_{xx}}{E_{xy}}\right).
\end{equation}

Non-equilibrium MD simulations for simple shear flows in the rotated Cartesian coordinates are performed in many MD cells according to the local strain rate $E'$'s defined at each lattice node of the CFD. 
Once a local stress tensor $P'_{\alpha\beta}$ is obtained at the MD level, the local stress at each lattice node $P_{\alpha\beta}$ in the original coordinate system is obtained by combining the pressure $p$ obtained a priori by CFD and a tensor $T'_{\alpha\beta}$ obtained by subtracting the isotropic normal stress components from $P'_{\alpha\beta}$ as
\begin{equation}\label{sec2_eq11}
P = \Theta^{T}[-p {\rm I} + T']\Theta
  =-p {\rm I} +\Theta^{T} T' \Theta,
\end{equation}
where I is the unit tensor. 

In the MD simulations, we solve the so-called SLLOD equations of motion with the Gaussian iso-kinetic thermostat:\cite{book:89AT, book:08EM, art:84EM}
\begin{subequations}\label{sec2_sllod}
\begin{equation}
\frac{d {\bm r}_j}{dt}=
\frac{{\bm p}_j}{m}+\dot \gamma {r_y}_j {\bm e}_x,
\end{equation}
\begin{equation}
\frac{d {\bm p}_j}{dt}=
{\bm f}_j - \dot \gamma {p_y}_j {\bm e}_x
-\zeta {\bm p}_j,
\end{equation}
\end{subequations}
where ${\bm e}_x$ is the unit vector in the $x$ direction and the index $j$ represents the $j$th particle  ($j=1,\cdots,N$).
${\bm r}_j$ and ${\bm p}_j + m\dot\gamma {r_y}_j{\bm e}_x$ are the position and momentum of the $j$th particle, respectively, 
${\bm f}_j$ is the force acting on the $j$th particle due to the LJ potential in Eq. (\ref{sec2_eq0}), and $\dot\gamma$ is the shear rate experienced by each MD cell, which is written as $\dot\gamma=2E'_{xy}$ in the present multiscale scheme.
Note that, in the SLLOD equations, ${\bm p}_j/m$ represents the deviation of velocity of each particle from the mean flow velocity $\dot\gamma {r_y}_j{\bm e}_x$ in the MD cell.
The friction coefficient $\zeta$ is determined to satisfy the constant temperature condition $dT/dt=0$ with $T=\sum_{j} |{\bm p}_j|^2/3mN$.
The friction coefficient $\zeta$ is calculated as
\begin{equation}\label{sec2_thermo}
\zeta=\sum_{j}({\bm f}_j\cdot{\bm p}_j-\dot\gamma {p_x}_j {p_y}_j)/\sum_{j}|{\bm p}_j|^2.
\end{equation}
We integrate Eq. (\ref{sec2_sllod}) with Eq. (\ref{sec2_thermo}) using the leapfrog algorithm\cite{art:84BC} with the Lees-Edwards sheared periodic boundary condition in the $y$ direction and periodic boundary conditions in each cubic MD cell.

The space integral of the instantaneous microscopic stress tensor reads as
\begin{align}\label{sec2_stress}
V P'_{\alpha\beta}=&
\frac{1}{m}\sum_{j=1}^N{p_\alpha}_j{p_\beta}_j
-\sum_{\rm all pairs} \frac{dU_{\rm LJ}(\xi)}{d\xi}\frac{\xi_\alpha\xi_\beta}{\xi}
\end{align}
where we rewrite the momentum of the $j$th particle, ${\bm p}_j + m\dot\gamma {r_y}_j{\bm e}_x$, as ${\bm p}_j$. 
Here, $V$ is the volume of the MD cell, i.e., $V=l_{\rm MD}^3$, and $\bm \xi$ is the relative vector ${\bm r}_j - {\bm r}_{j'}$ between the two particles ${\bm r}_j$ and ${\bm r}_{j'}$. 
The density is also fixed in each MD simulation by the number of particles and the box size. 
Thus, we calculate the local stress using so-called NVT-ensembles, $P'(\rho,T,E')$. 
The macroscopic stress is averaged in steady states after the transient behaviour has vanished. 
In the present computations, the transient time in each MD process is set as a tenth of the CFD time step, $\Delta t$/10. 
The initial states in each MD simulation are created by re-scaling the thermal velocities of molecules to be fixed to the local temperatures without changing the molecular configurations from those obtained by the previous process.

In the following, $\Delta t$ and $\Delta x$ represent the time-step and the mesh size of CFD calculations, and $t_{\rm MD}$ and $l_{\rm MD}$ represent the sampling duration and the side length of a MD cell, respectively. 
The two parameters $\Delta t/t_{\rm MD}$ and $\Delta x/l_{\rm MD}$ represent the efficiency of our multiscale simulations. 
In the present simulations, we fix the time step of the MD simulation $\Delta t_{\rm MD}$ as $\Delta t_{\rm MD}$=0.005. 
The sampling durations of the MD simulation $t_{\rm MD}$ are set to be larger than the correlation time of the temporal shear stress for the bulk fluids, $t_{\rm MD} \gg \tau_{LJ}$. 
The time step of CFD $\Delta t$ is set to be small enough for the CFD solutions to be stable.

\subsection{Driven cavity flows}
The simple LJ liquid is contained in a square box with a side length $L$. 
At $t$=0, the upper wall starts to move from left to right at a velocity $V_w$. 
A non-slip boundary condition is applied at each wall: $v_x$=$V_w$ and $v_y$=0 at $y$=$L$ and $v_x$=$v_y$=0 at the other walls. 
At the upper left and right corners, $v_x$=$V_w$ and $v_y$=0 are applied. 
Although the non-slip condition causes singularities in the strain rates and stresses at the upper corners in the continuum description, we use the conventional non-slip boundary condition even at the corners to compare the results to those obtained by the usual CFD calculation.\cite{art:95KB}
We note that multiscale simulations to resolve the singular forces at the corners can be found in Refs. \onlinecite{art:04NCR,art:06NRC}. The multiscale simulations are performed for the parameters listed in Table \ref{sec2_t1}. 
Here, the Reynolds number is defined as $\rho U L /\eta$, and the viscosity of the corresponding LJ liquid is $\eta$=1.7, which is calculated by the integral of the stress-relaxation function $G(t)$ shown in Fig. \ref{sec2_fig_gt}, $\eta=\int_0^\infty G(t) dt$. 
The computational domain is divided into 32$\times$32 uniform lattices.
\begin{table}[tbp]
\begin{center}
\begin{tabular}{c cccc cc cc cc cc cc cc}
\hline\hline
    && $l_{\rm MD}$
    && $t_{\rm MD}$
    && $N$
    && $L$
    && $U$
    && $Re$
    && $\Delta x/l_{\rm MD}$
    && $\Delta t/t_{\rm MD}$\\
    \hline
C1  && 6.84 && 4.68 && 256 && 219 && 0.46 && 59  && 1.0   && 1.0 \\
C2  && 6.84 && 9.36 && 256 && 438 && 0.91 && 235 && 2.0   && 2.0 \\
C3  && 6.84 && 3.11 && 256 && 876 && 1.83 && 941 && 4.0   && 4.0 \\
C4  && 8.62 && 9.36 && 512 && 438 && 0.91 && 235 && 1.59  && 2.0 \\
C5  && 8.62 && 18.7 && 512 && 438 && 0.91 && 235 && 1.59  && 1.0 \\
C6  && 10.86&& 15.7 &&1024 &&2780 && 0.58 && 941 && 8.0   && 8.0 \\
\hline\hline
\end{tabular}
\caption{
Simulation parameters for cavity flows.
$l_{\rm MD}$, $t_{\rm MD}$, and $N$ are the side length of MD cell, sampling duration of MD simulation, and number of particles contained in each MD cell, respectively. $L$, $U$, and $Re$ are the system size, velocity of the upper wall, and Reynolds number on the system, respectively.
$\Delta x/l_{\rm MD}$ and $\Delta t/t_{\rm MD}$ are the efficiency factors of the present multiscale simulations.
}
\label{sec2_t1}
\end{center}
\end{table}

\begin{figure*}[tbp]
\begin{center}
\includegraphics[scale=1]{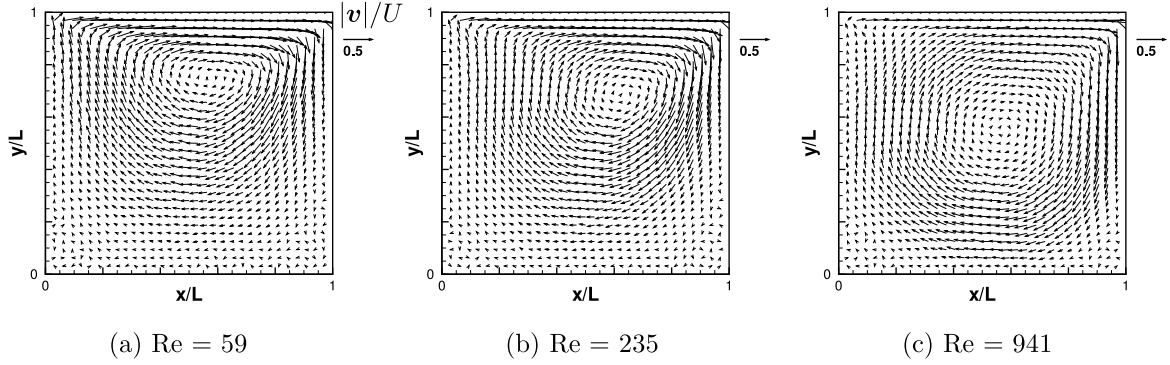}
\end{center}
\caption{The steady-state velocity profiles of the cavity flows for (a) Re=59 (C1 in Table \ref{sec2_t1}), (b) Re=235 (C2 in Table \ref{sec2_t1}), and (c) Re=941 (C3 in Table \ref{sec2_t1}).The velocity profiles are time averaged over $t/\Delta t$=[900,1000].}
\label{sec2_f4}
\end{figure*}

\begin{figure*}[tbp]
\begin{center}
\includegraphics[scale=0.9]{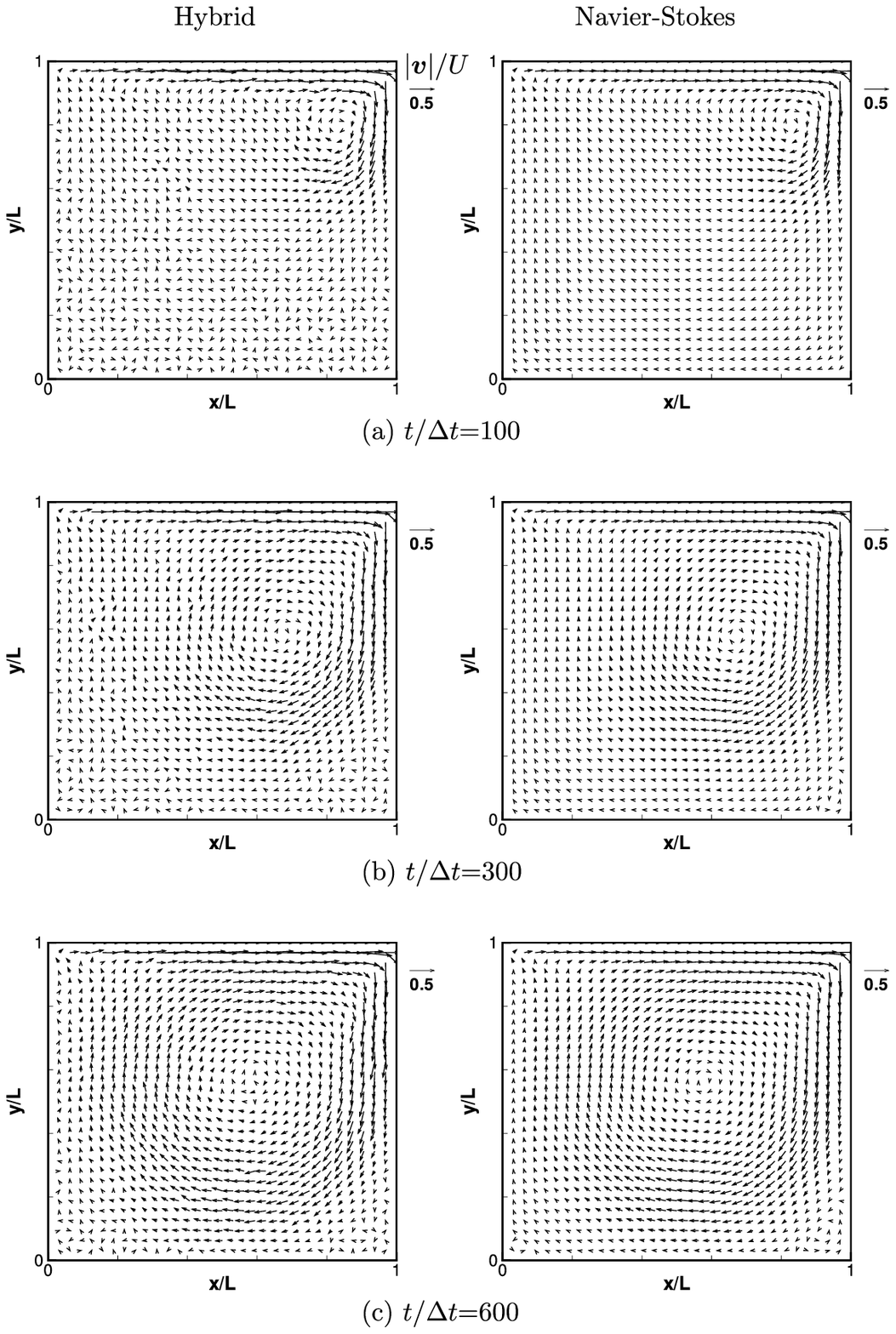}
\end{center}
\caption{Time evolutions of the velocity profile for the cavity flow with Re=941. 
The left column shows the results of the multiscale simulation with $\Delta x/l_{\rm MD}=\Delta t/t_{\rm MD}=8$ (C6 in Table \ref{sec2_t1}) and the right column shows the corresponding CFD results.
}\label{sec2_f5}
\end{figure*}

The results of the multiscale simulations are shown in Figs.~\ref{sec2_f4} and \ref{sec2_f5}. 
Figure \ref{sec2_f4} shows the steady-state velocity profiles time-averaged over $t/\Delta t$=[900,1000]. 
It is clear that our multiscale method can successfully reproduce the characteristic flow properties of cavity flows with different Reynolds numbers; the size of the vortex becomes larger as the Reynolds number increases. 
Figure \ref{sec2_f5} shows the snapshots of velocity profiles for the case of Re=980 at different time steps. 
Here, the results obtained by multiscale simulations with the efficiency factors $\Delta x/l_{\rm MD}$=$\Delta t/t_{\rm MD}$=8 are compared with the results obtained for the Newtonian fluid. 
The results show that a small vortex first appears in the upper-right corner and then moves gradually toward the centre of the box, with the size of the vortex increasing as time passes. 
Although fluctuations are seen in the instantaneous velocity profiles, our multiscale method can successfully reproduce the characteristic behaviours in the time-evolution of driven cavity flow.

\begin{figure*}[tbp]
\includegraphics[scale=1]{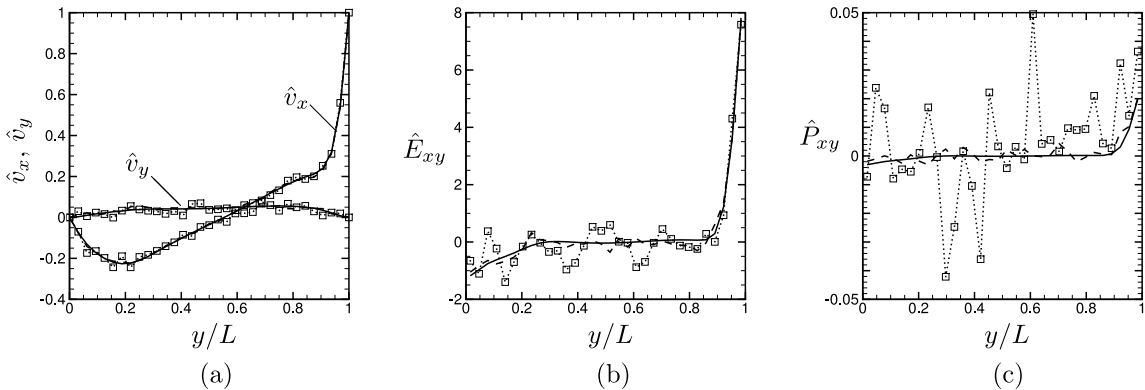}
\caption{
Comparisons between the instantaneous profiles of velocity $U \hat v_\alpha$, strain rate $(U/L) \hat E_{xy}$, and shear stress $(\rho U^2) \hat P_{xy}$ on the line of $x/L$=0.5 obtained by the multiscale simulation with $\Delta x/l_{\rm MD}=\Delta t/t_{\rm MD}=8$ (C6 in Table \ref{sec2_t1}) and those obtained by CFD for the cavity flow for $Re$=941.
The symbols show the results of the multiscale simulation and the solid lines those of CFD. The dashed lines in (b) and (c) show the time-averages over $t/\Delta t$=[900,1000].
}\label{sec2_1d_y_compari}
\end{figure*}

\begin{figure*}[tbp]
\includegraphics[scale=1]{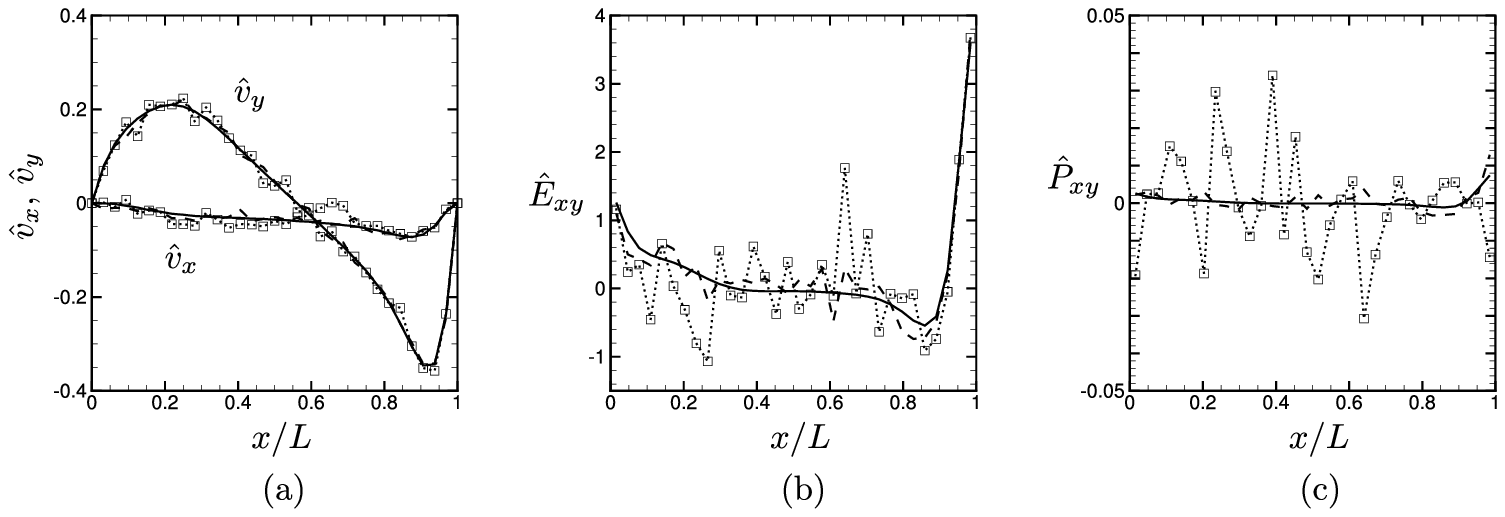}
\caption{
Comparisons of the instantaneous profiles of velocity $U \hat v_\alpha$, strain rate $(U/L) \hat E_{xy}$, and shear stress $(\rho U^2) \hat P_{xy}$ on the line of $y/L$=0.5 obtained by the multiscale simulation with $\Delta x/l_{\rm MD}=\Delta t/t_{\rm MD}=8$ (C6 in Table \ref{sec2_t1}) and those obtained by CFD for the cavity flow for $Re$=941. 
See also the caption in Fig. \ref{sec2_1d_y_compari}.
}\label{sec2_1d_x_compari}
\end{figure*}

The comparisons of the spatial variations of local velocities, strain rates, and stresses obtained by the multiscale simulation and those of the Newtonian fluid are shown in Figs. \ref{sec2_1d_y_compari} and \ref{sec2_1d_x_compari}. 
It is seen that the deviations of the local shear stresses obtained by the multiscale simulations are larger than those of the local velocities and strain rates. 
The instantaneous fluctuations of the local shear stresses are notable, but they are smoothed by taking the time averages. 
The fractional RMS deviations of the local velocities and stresses obtained by the multiscale simulations from those obtained by the usual CFD are also shown in Table \ref{sec2_t2}. 
The comparisons between C2, C4, and C5 show the effect of the efficiency factors $\Delta t/t_{\rm MD}$ and $\Delta x/l_{\rm MD}$ on the RMS deviations with fixed system parameters $L$, $U$, and $Re$. 
The RMS deviations increase as the efficiency factors increase with fixed system parameters, e.g., $L$, $U$, and $Re$. 
On the other hand, in the comparisons between C1 -- C3, the RMS deviations of the velocities slightly decrease in the order C1 -- C3, although the efficiency factors increase in the order C1--C3. 
This unexpected finding is explained by the fact that the local strain rates increase as the Reynolds number becomes larger, and thus the local shear stresses are also large in comparison to the noise intensity. 
The comparison of C3 and C6 also shows that the fractional RMS deviations decrease even though the efficiency factors increase. 
It should be noted that, in the simulation parameter C6, the number of particles contained in each MD cell is four times greater than in C3 and the sampling duration of the MD simulation is about five times longer than in C3. 
The values of C3 and C6 in Table \ref{sec2_t2} can be explained by the property of fluctuations discussed below. 
These facts indicate that multiscale simulations can be performed with high efficiency factors for large Reynolds numbers and large system sizes.
\begin{table}[tbp]
\begin{center}
\begin{tabular}{c cc cc}
\hline\hline
& \multicolumn{4}{c}{RMS deviation}\\
\cline{2-5}
& \multicolumn{2}{c}{${\bf v}$}&\multicolumn{2}{c}{$P_{xy}$}\\
\hline
C1 &&0.049 &&0.36 \\
C2 &&0.034 &&0.25 \\
C3 &&0.024 &&0.39 \\
C4 &&0.024 &&0.17 \\
C5 &&0.017 &&0.12 \\
C6 &&0.022 &&0.36 \\
\hline\hline
\end{tabular}
\end{center}
\caption{
Fractional root-mean-square (RMS) deviations in the steady states, which are defined as
$\sqrt{\int_0^Ldx^2\int_{\cal T}dt ( Q-Q_{\rm NS})^2/L^2{\cal T}}/{\rm Max}|Q_{\rm NS}|$,
for the velocity ($Q$=${\bm v}$) and for the shear stress ($Q$=$P_{xy}$). 
Here, ${\cal T}$ represents the sampling duration in the calculation of the RMS in the steady state.
}
\label{sec2_t2}
\end{table}

As mentioned above, the ratios $\Delta t/t_{\rm MD}$ and $\Delta x/l_{\rm MD}$ measure the efficiency of our multiscale simulations. 
For example, in the case of $\Delta t/t_{\rm MD}$=$\Delta x/l_{\rm MD}$=8, the computational efficiency is approximately $8^2\times 8$ times better than that of a full MD simulation. 
The larger the ratios, the more efficient the simulations; however, the statistical fluctuations also increase. 
In the following part, we will discuss the nature of the fluctuations in more detail.\cite{book:59LL,book:07K}
To handle the statistical noise explicitly, we rewrite Eq. (\ref{sec2_eq11}) as
\begin{equation}\label{sec2_eq13}
P=-pI+\Theta^T(T_*'+R')\Theta,
\end{equation}
where the off-diagonal stress tensor $T'$, which is to be determined by MD sampling, is decomposed into the non-fluctuating stress $T'_*$ and the fluctuating random stress $R'$ due to the thermal noise. 
The magnitude of each component of the random stress included in MD sampling, $\langle R_{{\rm MD}pq}'^2\rangle$ where $p$ and $q$ represent the index in Cartesian coordinates and do not follow the summation convention, should depend both on the size of the MD cell $l_{\rm MD}$ and the length of time $t_{\rm MD}$ over which the average is taken at the MD level; 
$\langle R_{{\rm MD}pq}'^2\rangle$=$\langle \bar R_{pq}(l_{\rm MD},t_{\rm MD})^2\rangle$, where $\bar R(l,t)$ represents the random stress tensor averaged in a cubic with a side length $l$ and over a time duration $t$.

At the CFD level, which is discretised with a mesh size $\Delta x$ and a time-step $\Delta t$, the physically correct magnitude should be $\langle R_{{\rm CFD}{pq}}'^2\rangle$=$\langle \bar R_{pq}(\Delta x,\Delta t)^2\rangle$. 
If the central-limit theorem, $\langle\bar R_{pq}(l,t)^2\rangle$$\propto 1/l^3 t$, is assumed, the following simple formula can be used:
\begin{equation}\label{sec2_eq14}
\langle R_{{\rm MD}{pq}}'^2\rangle=
\left( \frac{\Delta x}{l_{\rm MD}} \right)^3
\left(\frac{\Delta t}{t_{\rm MD}} \right)
\langle R_{{\rm CFD}{pq}}'^2\rangle.
\end{equation}
This line of reasoning leads to the following expression for the correctly fluctuating stress tensor $P$:
\begin{equation}\label{sec2_eq15}
P = -pI+\Theta^T\left[
T'_*+ 
\sqrt{
\left(\frac{l_{\rm MD}}{\Delta x}\right)^3
\left(\frac{t_{\rm MD}}{\Delta t}\right)
}
R'_{\rm MD}
\right] \Theta
\end{equation}
to be used in CFD instead of Eq.~(\ref{sec2_eq11}). 
This equation indicates that if we could re-weight the randomly fluctuating part ${R}'$ while the non-fluctuating part $T'_*$ remains untouched, hydrodynamic simulations including correct thermal fluctuations can be performed within the present framework. 
Incidentally, we can explain the results of the fractional RMS deviations in Table \ref{sec2_t2} using the relation (\ref{sec2_eq14}) while noting that the quantity $R$ is normalised by $\rho U^2$.

We note that the important key toward the development of the multiscale simulation is the separation of $T'_*$ and ${R}'$. 
We thus carried out the spectral analysis for the fluctuations in the total stress tensor computed directly from MD simulations, $T'=T'_* + {R}'$. 
The discrete Fourier transformation of $T'_{xy}$ is defined as
\begin{equation}
\varPi'_{xy}\{\bm k\}=\frac{1}{4M^2}\sum_{n_x=0}^{2M-1} \sum_{n_y=0}^{2M-1} \hat T'_{xy}\{\bm x\}\exp(-i{\bm k}\cdot{\bm x}),
\end{equation}
where ${\bm x}=(n_x\Delta x, n_y\Delta x)$ is the position of each lattice node $(n_x,n_y)$, ${\bm k}=(2\pi m_x/L, 2\pi m_y/L)$ is the wave vector, $n_x,n_y,m_z,m_y$ are integers, M is the lattice number in each $x$- and $y$-axis, and $\hat T'_{xy}\{ {\bm x} \}$ is defined as $\hat T'_{xy}\{ {\bm x} \}$=$T'_{xy}(x+\Delta x/2,y+\Delta x/2)$ for $0\le x, y \le L$, $\hat T'_{xy}\{ {\bm x} \}$=$T'_{xy}\{2L-x,y\}$ for $L<x\le 2L$, and $\hat T'_{xy}\{ {\bm x} \}$=$T'_{xy}\{x,2L-y\}$ for $L<y\le 2L$.

\begin{figure*}[tb]
\begin{center}
\includegraphics[scale=1]{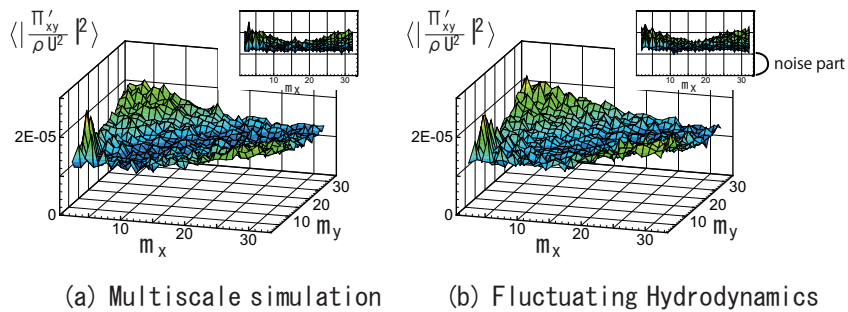}
\end{center}
\caption{
The fluctuations of $T'_{xy}$ for the case of cavity flow with Re=59. The power spectrum $\langle |\varPi'_{xy}\{\bm k\}|^2\rangle$ for the present multi-scale model with $\Delta x/l_{\rm MD}=\Delta t/t_{\rm MD}=1$ is shown in (a); the corresponding result from the fluctuating hydrodynamics is shown in (b) for comparison. $\varPi'_{xy}$ represents the discrete Fourier transform of $T'_{xy}$. $m_\alpha$ is defined as $m_\alpha$=$(L/2\pi)k_\alpha$, where ${\bm k}$ is the wave vector. The insets on each figure show the $\langle|\varPi'_{xy}|^2\rangle$-$m_x$ plane.
}\label{sec2_f6}
\end{figure*}
\begin{figure*}[htbp]
\begin{center}
\includegraphics[scale=1]{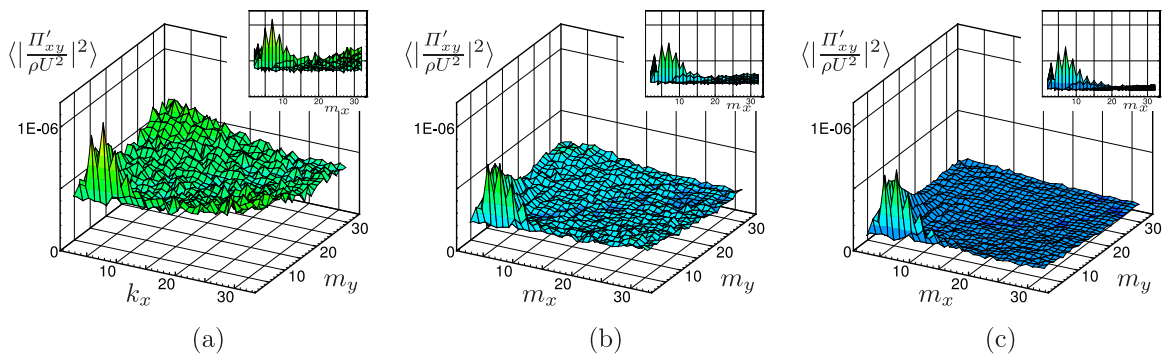}
\end{center}
\caption{
The fluctuations of $T'_{xy}$ for the case of cavity flow with Re=235. The power spectra $\langle |\varPi'_{xy}\{\bm k\}|^2\rangle$ is plotted in (a) for the case of Fig. \ref{sec2_f4} (b). In (b), only the number of particles is doubled; other parameters are unchanged from (a). In (c), both the number of particles and the sampling time of $T'_{xy}$ are doubled. $\varPi'_{xy}$ represents the discrete Fourier transform of $T'_{xy}$. $m_\alpha$ is defined as $m_\alpha$=$(L/2\pi)k_\alpha$, where ${\bm k}$ is the wave vector. The insets on each figure show the $\langle|\varPi'_{xy}|^2\rangle$-$m_x$ plane.
}
\label{sec2_f7}
\end{figure*}

The power spectra $\langle |\varPi'_{xy}\{\bm k\}|^2\rangle$ calculated from our multiscale simulations are plotted in Fig. \ref{sec2_f6} (a) for the case of $\Delta t/t_{\rm MD}=\Delta x/l_{\rm MD}=1$, which corresponds to the case of Fig. \ref{sec2_f4} (a) (C1 in Table \ref{sec2_t1}). 
The angle bracket $\langle\cdots\rangle$ indicates the time average taken at steady state at the CFD level. 
One can see that the overall structure is rather simple. 
There exist a relatively large peak around ${\bm k}=0$ and rather flat distributions throughout the ${\bm k}$ plane. 
The former corresponds to the contributions from the non-fluctuating part $T'_*$, and the latter corresponds to the contributions from the random stress $R'$. 
The same quantity obtained by conventional fluctuating hydrodynamics using a constant Newtonian viscosity and a random stress, the intensity of which is determined by the fluctuation-dissipation theorem \cite{book:59LL}, is shown in Fig. \ref{sec2_f6} (b) for comparison.\cite{note1}
Those two plots are surprisingly similar , including the fluctuation part. 
This similarity means that our multiscale simulation generates fluctuations quite consistent with fluctuating hydrodynamics using the fluctuation-dissipation theorem in the case of $\Delta x/l_{\rm MD}$=$\Delta t/t_{\rm MD}$=1.

In Fig. \ref{sec2_f7}, one sees how the fluctuations depend on the ratios $\Delta x/l_{\rm MD}$ and $\Delta t/t_{\rm MD}$. 
Here, in comparison to the reference case (a) [$\Delta x/l_{\rm MD}$=$\Delta t/t_{\rm MD}$=2], the number of particles used in the MD simulations are doubled in the case of (b) [$\Delta x/l_{\rm MD}$=1.59, $\Delta t/t_{\rm MD}$=2], and both the number of particles and the sampling duration used to take the time average are doubled in the case of (c) [$\Delta x/l_{\rm MD}$=1.59, $\Delta t/t_{\rm MD}$=1]. 
The noise intensity decreases with decreasing ratios $\Delta x/l_{\rm MD}$ and $\Delta t/t_{\rm MD}$, consistent with the central-limit theorem Eq. (\ref{sec2_eq14}); i.e., the noise intensity in (b) is approximately half that in (a), and the intensity in (c) is about one quarter of that in (a).

From the overall properties, we can confirm that the fluctuations arising in the present multiscale simulations are white noise, which obeys the central-limit theorem, written as Eq. (\ref{sec2_eq14}) or (\ref{sec2_eq15}). 
The results also suggest that some white-noise-reduction algorithm, such as a low-pass filter, could be useful in the CFD-MD coupling processes.

\section{one-dimensional polymeric flows}
As we have seen, the relaxation time of the local stresses in simple fluids is very short; for the LJ liquid in the previous section, it is estimated as $\tau_{LJ}<0.1$ in the LJ unit time (Fig. \ref{sec2_fig_gt}). 
Thus, we can predict the local equilibrium state at each instant in the sampling of the local stresses for the simple fluid (Fig. \ref{sec2_f00}). 
For polymeric liquids, however, the relaxation time is much longer than that of simple fluids, and thus it happens to be larger than the characteristic time of the macroscopic flows. 
In Fig. \ref{sec2_fig_gt}, we show the stress relaxation of a model polymer melt, which is discussed in this section. In this case, we cannot predict the local equilibrium states within the time-step duration to resolve the macroscopic flow behaviours. 
Instead, we must consider the history of the slow dynamics of polymer chains in each fluid element, i.e., the ``memory effect''.

In this section, we consider the flow behaviours of a polymeric liquid confined in parallel plates; we assume that the flow direction is restricted to being parallel to the upper and lower plates and that the local macroscopic quantities, e.g., the flow velocities, strain rates, and stresses, are uniform along the flow direction. 
This assumption allows us to calculate the macroscopic quantities using the usual fixed-mesh system without tracing the advection of a fluid element that contains a memory of the configuration of polymer chains. We note that one must treat the convection of the memory along the stream lines in the general two- or three-dimensional flows. 
The extension to polymeric flows with the advection of the memory will be given in the next section.

\begin{figure*}[tb]
\includegraphics[scale=1]{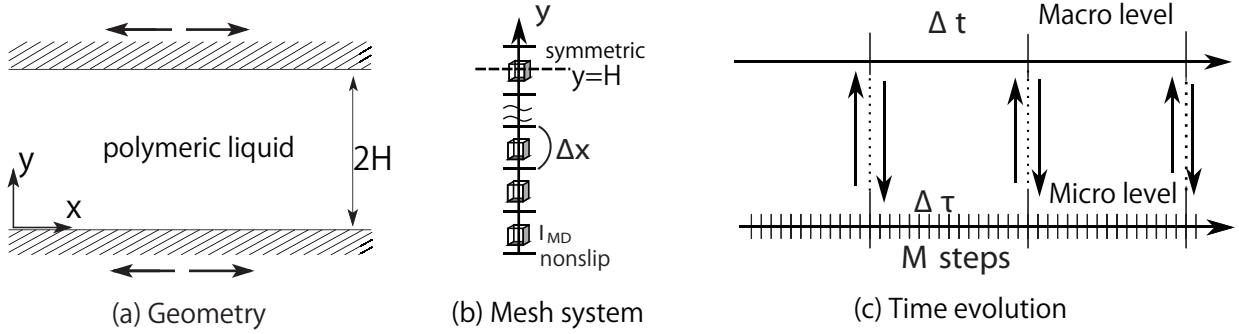}
\caption{
Schematics for the geometry of the problem, mesh system, and time-evolution scheme for one-dimensional flows of polymeric liquids in parallel plates.
}\label{sec3_f1}
\end{figure*}

\subsection{Multiscale modelling and the simulation method}
We consider a polymer melt with uniform density $\rho_0$ and temperature $T_0$ between two parallel plates (Fig. \ref{sec3_f1}(a)). 
Both plates can move in the $x$-direction. The melt is composed of short chains with ten beads. 
The number of beads comprising each chain in the MD simulation is represented by $N_{\rm b}$. 
Thus, $N_{\rm b}=10$. All of the beads interact with a truncated Lennard-Jones potential defined by Eq. (\ref{sec2_eq0}) in Sec. II.\cite{art:90KG} 
By using only the repulsive part of the Lennard-Jones potential, we prevent the spatial overlap of the particles. 
Consecutive beads on each chain are connected by an anharmonic spring potential
\begin{equation}\label{sec3_eq2}
U_{\rm F}(r)=-\frac{1}{2}k_c R_0^2 \ln
\left[
1-({r}/{R_0})^2
\right],
\end{equation}
where $k_c$=30$\epsilon/\sigma^2$ and $R_0$=$1.5\sigma$. 
The number density of the beads is fixed at $\rho_0/m$=1/$\sigma^3$, where $m$ is the mass of the bead particle. 
With this number density, the configuration of bead particles becomes severely jammed at low temperatures, resulting in the complicated non-Newtonian viscosity and long-time relaxation phenomena typically seen in glassy polymers.\cite{book:92M, book:96S, art:02YO}

We assume that the macroscopic quantities are uniform in the $x$- and $z$-directions, $\partial/\partial x$=$\partial/\partial z$=0. 
The macroscopic velocity $v_\alpha$ is described by the following equations:
\begin{equation}\label{sec3_eq3}
\rho_0\frac{\partial v_x}{\partial t} = 
\frac{\partial P_{xy}}{\partial y},
\end{equation}
and $v_y$=$v_z$=0, where $t$ is the time and $P_{\alpha\beta}$ is the stress tensor. 
Here and afterwards, the subscripts $\alpha$, $\beta$, and $\gamma$ represent the index in Cartesian coordinates, i.e., \{$\alpha$,$\beta$,$\gamma$\}=\{$x$,$y$,$z$\}. 
We also assume a non-slip boundary condition on each plate.

We use a usual finite volume method with a staggered arrangement for the CFD calculation; in this method, the velocity is computed at the node of each slit and the stress is computed at the centre of each slit. For the time-integration scheme, we use the simple explicit Euler method with a small time step $\Delta t$. A local stress is calculated at each time step of CFD using the non-equilibrium MD simulation with a small cubic MD cell with a side length $l_{\rm MD}$ associated with each slit according to a local strain rate. In each MD simulation, we solve the SLLOD equations of motion with the Gaussian iso-kinetic thermostat, Eq. (\ref{sec2_sllod}) with (\ref{sec2_thermo}), as in the previous section. The space integral of the microscopic stress tensor reads as
\begin{align}\label{sec3_eq6}
\sigma_{\alpha\beta}(t)=&
\frac{1}{m}\sum_{k=1}^{N_{\rm p}}\sum_{j=1}^{N_{\rm b}}{p_\alpha}^k_j{p_\beta}^k_j
-\sum_{\rm all pairs} \frac{dU_{\rm LJ}(\xi)}{d\xi}\frac{\xi_\alpha\xi_\beta}{\xi}
\nonumber\\
&-\sum_{k=1}^{N_{\rm p}}\sum_{j=1}^{N_{\rm b}}\frac{dU_{\rm F}(\xi)}{d\xi}\frac{\xi_\alpha\xi_\beta}{\xi},
\end{align}
where we rewrite the momentum of the $j$th bead on the $k$th chain, ${\bm p}^k_j + m\dot\gamma {r_y}^k_j{\bm e}_x$, as ${\bm p}_j^k$. Here, the indexes $k$ and $j$ represent the $k$th polymer chain ($k=1, \cdots, N_p$) and the $j$th bead ($j=1,\cdots,N_{\rm b}$) on each chain, respectively. $N_p$ and $N_b$ represent the number of polymer chains and beads on each chain, respectively. $\bm \xi$ in the right-hand side of Eq. (\ref{sec3_eq6}) represents the relative vector ${\bm r}^k_j - {\bm r}^{k'}_{j'}$ between the two beads, ${\bm r}^k_j$ and ${\bm r}^{k'}_{j'}$, in the second term and the relative vector ${\bm r}^k_j - {\bm r}^{k}_{j+1}$ between the two consecutive beads on the same chain, ${\bm r}^k_j$ and ${\bm r}^{k}_{j+1}$, in the third term.

In the present problem, we cannot assume a local equilibrium state at each time step of the CFD simulation because the relaxation time of the stress may become much longer than the time step of the CFD simulation (with which the macroscopic motions should be resolved). In the current simulations, the simple time averages of temporal stresses of the MD (averaged over the duration of a time-step of the CFD simulation) are used as the stresses at each time step of the CFD calculation without ignoring the transient time necessary for the MD system to be in the steady state (Fig. \ref{sec3_f1} (c)). Thus, the time integration of the macroscopic local stresses $\bar P_{\alpha\beta}$ is performed with the instantaneous microscopic stress tensor $\sigma_{xy}(t)$ as
\begin{align}\label{sec3_eq7}
\bar P_{\alpha\beta}(t,y)&=\int_t^{t+\Delta t} P_{\alpha\beta}(t',y)dt'
\nonumber\\
&=\frac{1}{l_{\rm MD}^3 }\int_{t}^{t+\Delta t}\sigma_{\alpha\beta}(\tau;\dot\gamma(t,y))d\tau,
\end{align}
where $\Delta t$ is the time step of the CFD simulation and $l_{\rm MD}$ is the side length of a cubic MD cell. 
The second argument of $\sigma_{\alpha\beta}$ in Eq. (\ref{sec3_eq7}), i.e., $\dot \gamma(t,y)$, is constant in the integral interval, which indicates that the shear rate to which each MD cell is subjected is also constant over a duration $\Delta t$ in each MD simulation. 
The final configuration of the molecules obtained at each MD cell at each time step of the CFD calculation is retained as the initial configuration for the MD cell at the next time step of the CFD. 
Thus, we trace all of the temporal evolutions of the microscopic configurations at the MD level so that the memory effects can be reproduced correctly. 
We can reduce the computation time needed for the spatial integration compared to that in a full MD simulation by using MD cells that are smaller than the slit size used in the CFD simulation. 
The performance efficiency of the present multiscale simulation is represented by a saving factor defined as the ratio of the slit size used in the CFD simulation $\Delta x$ to the cell size used in the MD simulation $l_{\rm MD}$, $\Delta x$/$l_{\rm MD}$. 
It also should be noted that, in addition to the saving factor $\Delta x/l_{\rm MD}$, the present multiscale method is quite suitable for a parallel computational algorithm because the MD simulations associated with each mesh of the CFD, which consume a large portion of the total simulation time, are performed independently.

We measure the space and time in the units of $\sigma$ and $\tau_0=\sqrt{m\sigma^2/\epsilon}$, respectively, just as in the previous section. 
The temperature $T$ is measured in the unit of $\epsilon/k_B$. 
In the following simulations, the temperature $T$ and density $\rho$ of the melt are uniform and fixed as $T=0.2$ and $\rho=1.0$. 
At this number density and low temperature, the polymer melt involves complicated non-Newtonian rheology. 
The time step of the CFD simulation $\Delta t$, the sampling duration of the MD simulation $t_{\rm MD}$, and the time step of the MD simulation $\Delta \tau$ are fixed as $\Delta t$=$t_{\rm MD}$=1 and $\Delta \tau$=0.001. 
Thus, 1000 MD steps ($M=1000$ in Fig. \ref{sec3_f1}.) are performed in each MD cell at each time step of the CFD computation. 
One hundred chains with ten beads are confined in each cubic MD cell with a side length $l_{\rm MD}$=10; thus, $N_{\rm p}$=100 and $N_{\rm b}$=10.

\subsection{Oscillatory flows}
The lower and upper plates start to oscillate with a constant angular frequency $\omega_0$ at a time $t=0$ as, respectively,
\begin{equation}\label{sec3_eq8}
v_w = \pm v_0 \cos(\omega_0 t),
\end{equation}
with $v_0=\Gamma_0 \omega_0 H$. 
Here, $\Gamma_0$ represents the amplitude of strain on the system. In the following simulations, the width between the plates, 2$H$, is fixed as 2$H$=5000. 
In the CFD calculations, the lower half of the volume between the plates is divided into 128 (or 64) mesh intervals with a mesh size of $\Delta x$=19.5 (or 39.0), and the symmetric condition is used at the middle between the plates. 
The 128-mesh system is used for a large $\Gamma_0$, and the 64-mesh system is used for a small $\Gamma_0$. 
Thus, the saving factor $\Delta x/l_{\rm MD}$ in the multiscale simulations is $\Delta x/l_{\rm MD}$=1.95 (or 3.9).

When the velocity amplitude of the oscillating plates $v_0$ is small enough, $v_0\ll 1$ (or $\Gamma_0\ll 1$), one can assume the linear response of local stress to the local strain rate as
\begin{equation}\label{sec3_eq9}
P_{xy}=\int_{-\infty}^t G(t-t')\dot\gamma(t')dt',
\end{equation}
where $G(t)$ is the stress-relaxation function, which is shown in Fig. \ref{sec2_fig_gt}. By taking the Fourier transform of Eq. (\ref{sec3_eq3}) with Eq. (\ref{sec3_eq9}), we obtain
\begin{equation}\label{sec3_eq10}
\frac{d^2 v_x^\omega}{dy^2}=\frac{{\rm i}\rho_0\omega}{\eta^*}v_x^\omega,
\end{equation}
where $v_x^\omega$ is the Fourier transform of the local velocity, $v_x^\omega(y)=\int_{-\infty}^{\infty}v(t,y)e^{-{\rm i}\omega t}dt$, and $\eta^*$ is the complex viscosity, $\eta^*=\eta'+{\rm i}\eta''$, obtained by the Fourier transform of the stress-relaxation function as $\eta^*(\omega)=\int_0^\infty G(t) e^{-{\rm i}\omega t}dt$. Together with the boundary condition, Eq. (\ref{sec3_eq8}), the analytical solution for the linear response regime is
\begin{widetext}
\begin{equation}\label{sec3_eq11}
v_x(y,t)=v_0
\left\{
e^{-k_0'y}\exp[{\rm i}(\omega_0t-k_0''y)]
-e^{-k_0'(2H-y)}\exp[{\rm i}(\omega_0t-k_0''(2H-y))]
\right\}
\left(
1-e^{-2 k_0^* H}
\right)^{-1},
\end{equation}
\end{widetext}
with 
\begin{subequations}\label{sec3_eq12}
\begin{align}
k_0'&=\sqrt{\frac{\rho_0\omega_0}{2|\eta^*(\omega_0)|}}
\left(
\cos\frac{\theta}{2}
-\sin\frac{\theta}{2}
\right),
\\
k_0''&=\sqrt{\frac{\rho_0\omega_0}{2|\eta^*(\omega_0)|}}
\left(
\cos\frac{\theta}{2}
+\sin\frac{\theta}{2}
\right),
\end{align}
\end{subequations}
and $k_0^*\equiv k_0'+{\rm i}k_0''$, where $\theta$ is the loss tangent defined as $\theta=\tan^{-1}(\eta''/\eta')$. 
The complex viscosities $\eta'$ and $\eta''$ can be calculated from the numerical data of the stress-relaxation function $G(t)$ shown in Fig. {\ref{sec2_fig_gt}. 
In the present paper, the analytical solution of Eq. (\ref{sec3_eq11}) with Eq. (\ref{sec3_eq12}) is used to show the validation of our multiscale simulations. 
We will demonstrate that the results of the multiscale simulation become closer to the analytical solutions as the strain amplitude $\Gamma_0$ decreases.

\begin{figure*}[htbp]
\includegraphics[scale=1]{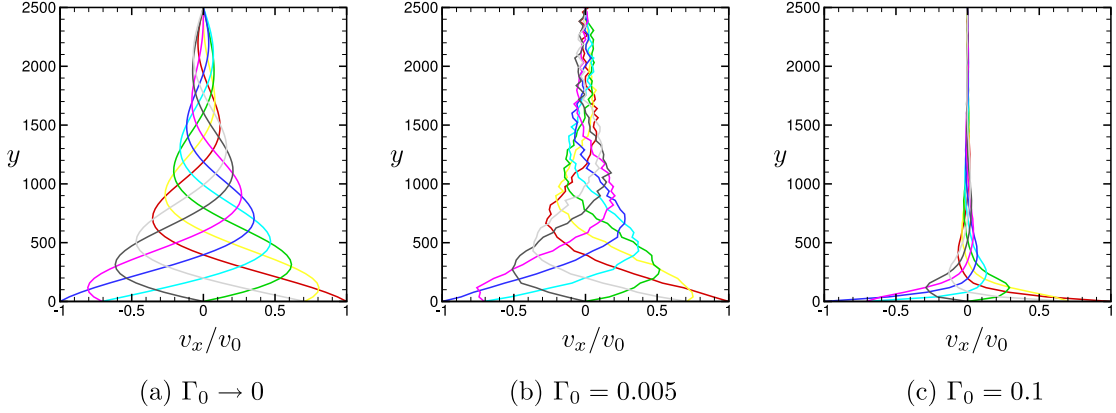}
\caption{
Velocity profiles of the melt at each $\pi/4$ phase shift for $\omega_0=6.1\times 10^{-3}$.
(a) The analytical solution with the numerical data of $\eta'$ and $\eta''$ calculated by the Fourier transform of $G(t)$ shown in Fig. \ref{sec2_fig_gt}.
(b) and (c) The results of the multiscale simulations for $\Gamma_0$=0.005 and 0.1, respectively, where instantaneous velocity profiles at each fixed phase are averaged in the steady oscillation.
}\label{sec3_f3}
\end{figure*}
\begin{figure*}[htbp]
\includegraphics[scale=1]{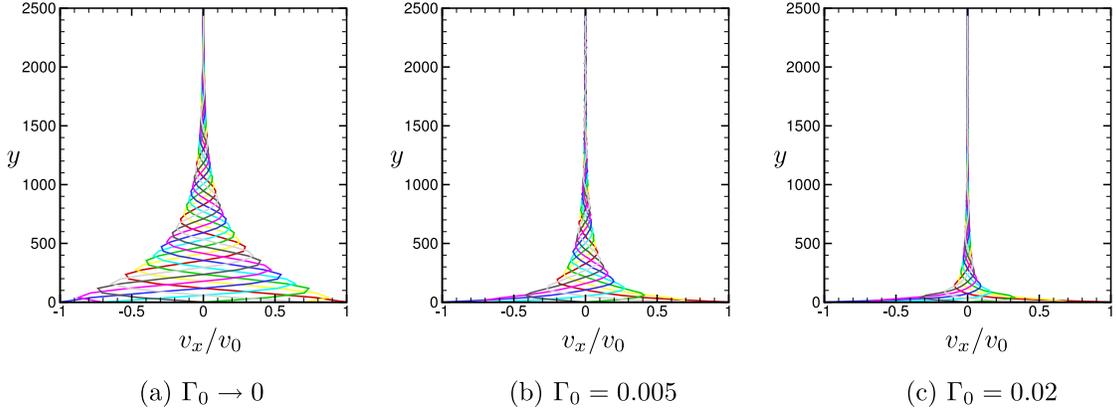}
\caption{
Velocity profiles of the melt at each $\pi/4$ phase shift for $\omega_0=0.025$.
See also the caption in Fig. \ref{sec3_f3}.
}\label{sec3_f4}
\end{figure*}

\begin{figure*}[htbp]
\includegraphics[scale=1]{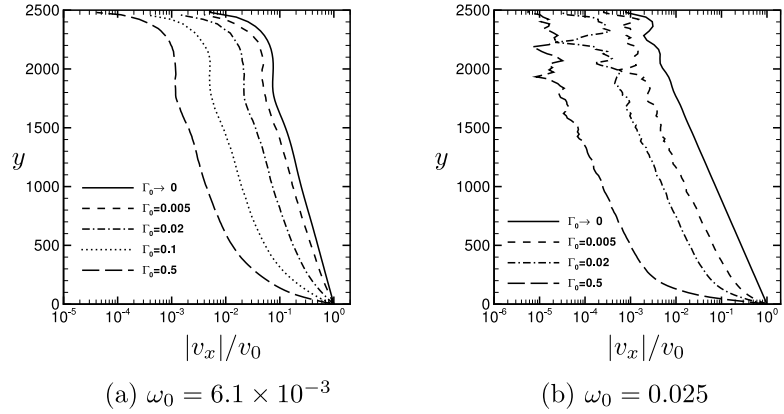}
\caption{
Comparison of the amplitudes of local velocity $|v_x|$ for various strain amplitudes on the system $\Gamma_0$.
(a) $\omega_0=6.1\times 10^{-3}$ and (b) $\omega_0=0.025$.
}\label{sec3_f5}
\end{figure*}

\begin{figure}[htbp]
\includegraphics[scale=1]{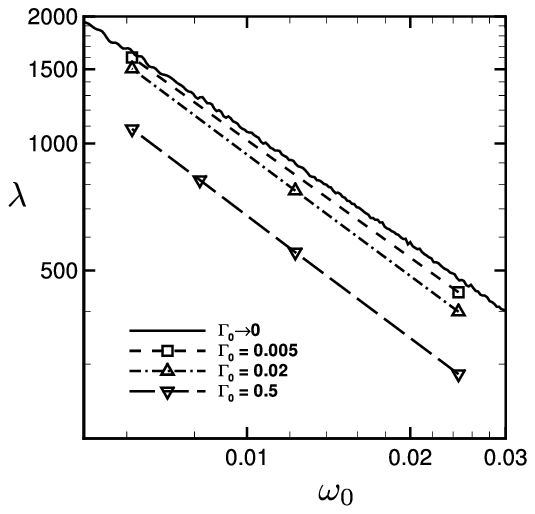}
\caption{
Comparison of the wavelengths $\lambda$ for various strain amplitudes on the system $\Gamma_0$.
}\label{sec3_f6}
\end{figure}

Figures \ref{sec3_f3} and \ref{sec3_f4} show the velocity profiles of the polymer melt in steady oscillations. 
It is seen that, as the amplitude of the oscillating plates $\Gamma_0$ increases, the thickness of the boundary layer becomes thinner. 
This phenomenon is related to the shear thinning occurring near the oscillating plate, where the local strains are large enough for the local viscosities to deviate from those in the linear-response regime.

With use of the Fourier transform for the time evolutions of local velocities and selecting the mode of the oscillation frequency of the plate $\omega_0$, we can express the time evolution of the oscillation velocity as
\begin{equation}\label{sec3_eq13}
v_x(y,t)=|v_x|(y)\cos(\omega t + \phi(y)).
\end{equation}
Here, the contributions of odd higher harmonics, 3$\omega_0$, 5$\omega_0$, and $\cdots$, are ignored because they are much smaller than the mode of $\omega_0$.\cite{art:10YY}
A detailed investigation of the nonlinear rheology of the present melt can be found in Ref. \onlinecite{art:10YYb}. 
Figure \ref{sec3_f5} shows the comparison of the amplitudes of local velocity $|v_x|$ between the various amplitudes of strain on the system $\Gamma_0$. 
As the amplitude of oscillating plates $\Gamma_0$ decreases, the profiles of the amplitudes of the local velocity $|v_x|$ become closer to that for the linear response. 
At large values of $\Gamma_0$, $|v_x|$ rapidly decreases near the oscillating plate and approaches the linear-response profile as the distance from the plate increases. 
One can also measure the wavelength for the velocity profile, $\lambda$, using the relation $\phi(\lambda)=0$, where $\phi(y)$ is the local phase retardation in Eq. (\ref{sec3_eq13}). 
For the analytical solution Eq. (\ref{sec3_eq11}), the wavelength $\lambda$ is written as $\lambda=2\pi/k_0''$. Figure \ref{sec3_f6} shows the comparisons of the wavelengths $\lambda$ between various strain amplitudes for the system $\Gamma_0$. 
It is seen that the wavelengths are very close to those for the linear response at all oscillation frequencies $\omega_0$ at $\Gamma_0$=0.005 but that they deviate more strongly from those for the linear-response regime as $\Gamma_0$ increases. 
The deviation is slightly larger because the oscillation frequency $\omega_0$ is larger at the same value of $\Gamma_0$.

\begin{figure*}[tbp]
\includegraphics[scale=1.0]{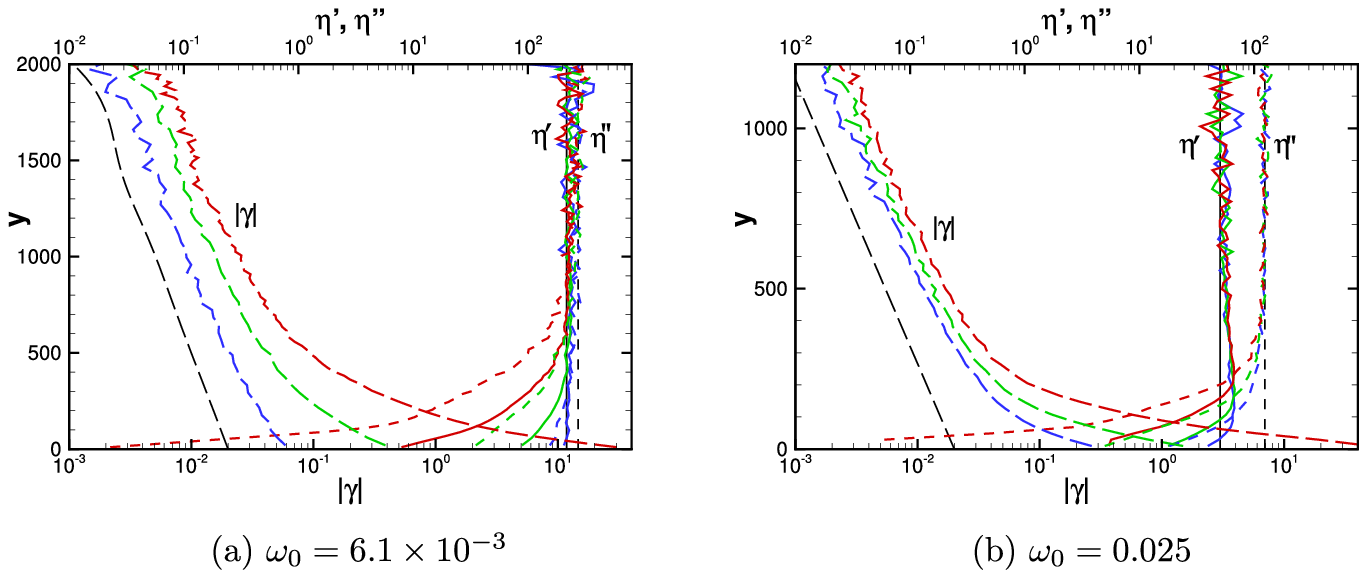}
\caption{
Spatial variations of the local complex viscosities $\eta'$ (solid line) and $\eta''$ (dashed line) and the amplitude of local strain $|\gamma|$ (long-dashed line) for the oscillation frequencies $\omega_0=6.1\times 10^{-3}$ (a) and 0.025 (b).
The blue, green, and red colours show the results for $\Gamma_0$=0.005, 0.02, and 0.5, respectively, in both figures.
The black lines are the results for the linear response for infinitesimally small strains.
}\label{sec3_f7}
\end{figure*}

We also measure the ``local'' viscoelastic properties in terms of the ``local'' complex viscosity $\eta^*$(=$\eta'+{\rm i}\eta''$). 
The real and imaginary parts of the complex viscosity, $\eta'$ and $\eta''$, represent the viscous and elastic responses, respectively, in the shear stress. 
The complex viscosities $\eta'$ and $\eta''$ can be calculated as follows. 
As in the case of the oscillation velocity, by using the Fourier transform of the time evolution of the local strain rates and selecting the mode of the oscillation frequency $\omega_0$, we can reconstruct the time evolution of the local strain rate $\dot\gamma$ in the form of $\dot\gamma=|\dot\gamma|(y)\cos(\omega_0 t + \psi(y))$. 
In the same way, we can also write the local shear stresses as $P_{xy}=P_{xy}'(y)\cos(\omega_0 t\psi)-P_{xy}''(y)\cos(\omega_0 t\psi)$. 
The complex viscosities $\eta'$ and $\eta''$ are obtained by $\eta'=P_{xy}'/|\dot\gamma|$ and $\eta''=P_{xy}''/|\dot\gamma|$, respectively.\cite{art:10YY}

Figure \ref{sec3_f7} shows the spatial variations of the local complex viscosities $\eta'$ and $\eta''$ and the amplitude of the local strain $|\gamma|$ (=$|\dot\gamma|/\omega_0$) for the oscillation frequencies $\omega_0=6.1\times 10^{-3}$ and 0.025 for various strain amplitudes on the system $\Gamma_0$. 
For comparison, the results for the linear response are also shown. 
As the distance from the plate increases, the amplitude of the local strain $|\gamma |$ decreases monotonically. 
When $|\gamma |$ is smaller than a few percent, the local complex viscosities become constant, and these constant values correspond to those for the linear response. 
As the amplitude of the oscillating plates decreases, the profile of the amplitude of local strain $|\gamma|$ approaches that for the linear response. 
In addition, the range in which the local complex viscosities agree with those for the linear response broadens near the plate. 
In particular, the dynamic viscosity $\eta'$ in Fig. \ref{sec3_f7}(a) is almost uniform over the entire domain at $\Gamma_0=0.005$. Shear-thinning is observed in the vicinity of the oscillating plate. 
That is, the complex viscosities $\eta'$ and $\eta''$ decrease near the oscillating plate, where the local strain rapidly increases. 
The elasticity $\eta''$ decreases more rapidly near the oscillating plate than the viscosity $\eta'$; thus, the viscosity $\eta'$ is dominant in close vicinity to the plate, while the elasticity $\eta''$ is larger than the viscosity $\eta'$ far from the plate. 
In the case of a high oscillation frequency and large amplitude of oscillating plates, say $\omega_0$=0.025 and $\Gamma_0=0.5$, the melt behaves as a viscous liquid in the vicinity of the plate because the viscosity $\eta'$ is much larger than $\eta''$, $\eta'\gg\eta''$, but as a viscoelastic solid far from the plate because the elasticity $\eta''$ is larger than $\eta'$, $\eta'>\eta''$. 
In the transient region, the melt behaves as a viscoelastic liquid. 
Thus, the melt has three different rheological regimes of viscous liquid, viscoelastic liquid, and viscoelastic solid over the oscillating plate at high oscillation frequencies and large amplitudes of the oscillating plate.

The crossover of the local rheological properties can be characterised by the local Deborah numbers $De$ defined by the oscillation frequency of the plate $\omega_0$ and the local relaxation times of the monomer structures and polymer-chain configurations, $\tau^\alpha$ and $\tau^R$ (which are referred to as the $\alpha$ relaxation time and the Rouse relaxation time, respectively), as $De^\alpha=\omega_0\tau^\alpha$ and $De^R=\omega_0\tau^R$. 
The relaxation times $\tau^\alpha$ and $\tau^R$ decrease as the strain rate $\dot\gamma$ increases. 
Thus, the local Deborah numbers are smaller closer to the oscillating plate. 
We found that the crossover of $\eta'$ and $\eta''$ occurs at the position where the local Deborah number $De^\alpha$ is unity. 
In the viscous liquid regime, the local Deborah number $De^R$ is less than unity. 
For more details, consult Refs. \onlinecite{art:09YY, art:10YY}.

\section{Advection of strain-rate memory in two-dimensional polymeric flow}
In the previous section,
we described ``the memory effect'' 
in one-dimensional polymeric flows.
In general two-dimensional polymeric flows, 
we must consider
the advection of the strain-rate memory
that is preserved in microscopic degrees of freedom in a fluid element.
Because the general two-dimensional flow is inhomogeneous 
along the flow direction, 
the advection of the memory affects its flow behaviour
when the characteristic relaxation time $\tau$ of a polymeric liquid
is larger than the transit time $\tau_{\rm t}$ 
that the polymeric liquid takes 
to move 
through a distance equal to
the size of a fluid element.
Polymers in the fluid element
experience a local strain rate during $\tau_{\rm t}$.
When a strain rate is imposed on the fluid element during
$\tau_{\rm t}$ ($< \tau$) in the inhomogeneous flow,
the stress response of the fluid element is different 
from that in the steady state under the same strain rate
because the strain-rate memory in the stress persists during $\tau$ 
longer than $\tau_{\rm t}$.
Therefore, the stress does not reach the value at the steady state.
Because the transit time becomes infinitesimal as we decrease 
the fluid-element size 
to increase the spatial resolution 
of the numerical simulation,
consideration of the advection of the memory is essential 
for a polymeric liquid
with a finite relaxation time
in inhomogeneous flow fields.

To deal with the advection of the memory
in a manner consistent with the macroscopic flow,
we employ a Lagrangian fluid method
in which microscopic simulators are 
assigned to the fluid elements.\cite{MT2010a} 
Because each fluid element moves along the flow direction, 
the microscopic variables are advected in a Lagrangian manner.

In this section, we discuss the flow of a well-entangled polymer melt 
that is inhomogeneous in the flow direction.
The polymer melt discussed in the previous section was
composed of short polymer chains without entanglements among polymers.
In the short-chain system, the memory effect is mainly caused by
the relaxation of the polymer orientation.
In the polymer melt flow without entanglements, 
the relaxation time of the polymer orientation can be considered to be
the same as the relaxation time of the polymer's extension
because the origins of these relaxations are same.
That is,
both relaxation times are determined by the competition between
the following two forces: i) an elastic force coming 
from the chain-conformational entropy
and 
ii) a frictional force coming from the interaction between beads.
In an entangled polymer melt, however,
these two characteristic relaxations 
are fundamentally different because of the presence of the entanglements.
Therefore, in dealing with entangled polymer melts,
we will observe the effects of each of these different relaxations 
on the flow behaviour.
We will ultimately conclude that 
the memory effect 
relates not only to the relaxation of orientation
but also to the relaxation of extension.\cite{MT2010b}

\subsection{Lagrangian fluid model with microscopic internal
  degrees of freedom}

We consider a polymer melt composed of well-entangled linear polymers.
In this polymer melt, the number of beads between two adjacent
entanglements along the polymer chain
has been found to be approximately 35 
in terms of the MD polymer chain.\cite{art:90KG}
Therefore, to describe the entangled polymer melt,
each polymer chain in the MD simulation must consist of more than
100 beads.
Dealing with the entangled polymer melt 
in the multiscale simulation with MD simulators
seems impossible
because of the extremely high numerical expense.
To decrease the numerical cost of the multiscale simulation,
we employ a coarse-grained slip-link model 
that reduces the 35 beads to a single primitive path. 
Specifically, by using the coarse-grained model,
the degrees of freedom become less than 1/35$^{\text{th}}$ -of those 
in the corresponding MD simulation.
The great reduction in the numerical expense 
using the coarse-grained model 
enables us to simulate the entangled polymer melt
in the multiscale approach with almost the 
same numerical expense as in the case of
the unentangled polymer melt with the MD simulation.

As in the previous sections,
we solve the macroscopic momentum equation of fluid
\begin{align}
\rho_i 
\derivative{\bol{v}_i}{t}
&=\bol{\nabla}\cdot\bol{\sigma}_i-\bol{\nabla}p_i
+ \bol{F}^{\rm b},\label{eq:velocity}
\end{align}
where the subscript
$i$ represents the $i$-th fluid element 
that is handled as a particle 
at the position $\bol{r}_i$,
$\rho_i$ is the density, $\bol{v}_i$ is the velocity vector, 
$\bol{\sigma}_i$ is the stress tensor,
$p_i$ is the isotropic pressure,
and
$\bol{F}^{\rm b}$ is a body force.
Moreover, we need to solve the time evolution of 
the position $\bol{r}_i$
that follows 
\begin{align}
\derivative{\bol{r}_i}{t}
&=\bol{v}_i \label{eq:position}
\end{align}
to consider the fluid
in a Lagrangian manner.
Once we interpret 
the momentum equation of the fluid (Eq. \eqref{eq:velocity}) 
as
\begin{align*}
\rho_i \derivative{\bol{v}_i}{t}
=\bol{F}_i, \tag{\ref{eq:velocity}'}
\end{align*}
where $\bol{F}_i$ is the right-hand side of Eq. \eqref{eq:velocity},
time-integral schemes 
developed in MD simulations are available.
Here we implement the velocity-Verlet algorithm to update $\{\bol{r}_i\}$ 
and $\{\bol{v}_i\}$.

\begin{figure}[t]
\begin{center}
\includegraphics[scale=0.5]{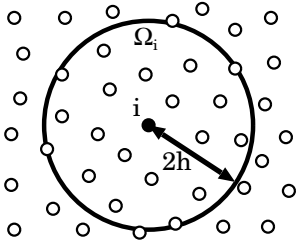}
\end{center}
\caption{\label{fig:sph}
Interaction circles with a cut-off length $2h$ in two-dimensional SPH.
The small filled and vacant circles ($\bullet,\circ$) represent fluid particles.
The large solid circle represents the interaction range
of the $i$-th fluid particle.}
\end{figure}

To solve Eq. \eqref{eq:velocity},
we need to know the derivatives of the stress and pressure
at the position $\bol{r}_i$, 
$(\bol{\nabla}\cdot\bol{\sigma})_i$ and $(\bol{\nabla}p)_i$.
Because we now know that 
the stress and pressure have values 
only at the positions $\{\bol{r}_i\}$, 
we cannot obtain their derivatives in a straightforward manner.
Instead of directly obtaining the derivatives 
$(\bol{\nabla}\cdot \bol{\sigma})_i$ and $(\bol{\nabla}p)_i$,
solving a linear matrix equation 
composed of interpolated values of $\bol{\sigma}_i$ and $p_i$
enables us to obtain their derivatives.
The procedures are explained in the following paragraphs.

In the Lagrangian fluid simulation,
as the positions of the fluid particles $\{\bol{r}_i\}$ 
change over time,
interpolation techniques
of physical variables at a certain position
between the fluid particles are required 
to obtain the spatial derivatives 
$\bol{\nabla}f$ or $\bol{\nabla}\cdot \bol{f}$ 
where $f$ ($\bol{f}$) indicates an arbitrary scalar variable (vector or tensor).
In the Eulerian fluid simulation,
as the interpolation scheme has been well established,
we do not need to consider
the interpolation scheme.
In our Lagrangian fluid dynamics simulation,
according to the smoothed particle hydrodynamics (SPH) simulation,\cite{SPH}
we employ the Gaussian kernel interpolation;
\begin{equation}
\tilde{f}(\bol{r}_i)
\equiv a^{d}\sum_{j \in \Omega_i} f(\bol{r}_j) W(|\bol{r}_i-\bol{r}_j|,h),
\label{eq:sph}
\end{equation}
where $f(\bol{r}_i)$ represents an arbitrary physical variable 
at the position $\bol{r}_i$ in the fluid,
$\tilde{f}(\bol{r}_i)$ is the interpolated variable of $f$,
$a$ is the unit length in the fluid particle simulation,
and $d$ is the dimensionality of the space.
The window function $W(x,h)$ is assumed to be
a Gaussian shaped function of distance $x$ and
half-value width $h=1.2a$, and it follows
\begin{align}
W(x,h)=\begin{cases}
\frac{A_d}{(h\sqrt{\pi})^d}
\left(e^{-\frac{x^2}{h^2}}-e^{-4}\right) & (|x| \leq 2h),\\
0 & (\mbox{otherwise}),
\end{cases}
\end{align}
where the cut-off radius is $2h$. 
The normalisation parameter $A_d$ equals 1.04823, 1.10081 and 1.18516
for $d=$ 1, 2 and 3, respectively.\cite{MSPH}
To obtain an interpolated variable $\tilde{f}$
of a certain variable $f$ at the position $\bol{r}_i$ of the $i$-th
fluid particle,
fluid particles in the circular (two-dimensional)/ spherical
(three-dimensional) region
$\Omega_i$ with radius $2h$ in which the $i$-th fluid particle is placed 
at the centre are accounted for, as shown in Fig. \ref{fig:sph}.
Hereafter we will use variables with tildes to represent interpolated values 
using Eq. \eqref{eq:sph}.

To obtain the spatial derivatives of $f(\bol{r}_i)$,
i.e., $\partial_{\alpha} f$ ($\alpha=\{x,y,z\}$) 
at the position $\bol{r}_i$,
we solve the following linear equation of the matrix form
(see Refs. \onlinecite{MT2010a,MSPH,FPM}):
\begin{widetext}
\begin{equation}
\begin{pmatrix}
\tilde{f}(\bol{r}_i)\\
\del_{\delta}\tilde{f}(\bol{r}_i)\\
\del_{\epsilon}\del_{\zeta}\tilde{f}(\bol{r}_i)
\end{pmatrix}=
\begin{pmatrix}
\tilde{1} & \tilde{R^{\alpha}} &
C^{\beta,\gamma} \tilde{R^{\beta}R^{\gamma}}\\
\del_{\delta}(\tilde{1}) & \del_{\delta}(\tilde{R^{\alpha}}) &
C^{\beta,\gamma}\del_{\delta}(\tilde{R^{\beta}R^{\gamma}}) \\
\del_{\epsilon}\del_{\zeta}(\tilde{1}) & 
\del_{\epsilon}\del_{\zeta}(\tilde{R^{\alpha}})&
C^{\beta,\gamma}\del_{\epsilon}\del_{\zeta}(\tilde{R^{\beta}R^{\gamma}}) \\
\end{pmatrix}
\begin{pmatrix}
f(\bol{r}_i)\\
\del_\alpha f(\bol{r}_i)\\
\del_{\beta}\del_{\gamma} f(\bol{r}_i)
\end{pmatrix}
,\label{eq:MSPH}
\end{equation}
\end{widetext}
where $C^{\beta,\gamma}\equiv 1-\frac{1}{2}\delta_{\beta,\gamma}$
and $\delta_{\beta,\gamma}$ is Kronecker's delta.
Here we used a reduced representation; 
Greek indexes represent the coordinate axes 
and the number of rows in Eq. \eqref{eq:MSPH} is seven in two dimensions
and 13 in three dimensions.
The derivatives of the interpolated variables 
in the left-hand side in Eq. \eqref{eq:MSPH} are defined as
\begin{align}
\del_{\alpha} \tilde{f}(\bol{r}_i) 
&\equiv a^d \sum_{j\in \Omega_i} f(\bol{r}_j) 
\del_{\alpha}W(|\bol{r}_i-\bol{r}_j|,h),\label{eq:msph1}\\
\del_{\alpha}\del_{\beta}\tilde{f}(\bol{r}_i) 
&\equiv a^d \sum_{j\in \Omega_i} f(\bol{r}_j) 
\del_{\alpha} \del_{\beta}W(|\bol{r}_i-\bol{r}_j|,h),\label{eq:msph2}
\end{align}
and the interpolated values of positional difference 
between the $i$-th and $j$-th 
fluid particles $\bol{R}\equiv \bol{r}_i-\bol{r}_j$
and dyadic $\bol{R}\bol{R}$ are, respectively,
\begin{align}
\tilde{R^{\alpha}}&
\equiv a^d\sum_{j\in \Omega_i}
 R^{\alpha}W(|\bol{r}_i-\bol{r}_j|,h),\label{eq:msph3}\\
\tilde{R^{\alpha}R^{\beta}}&
\equiv a^d\sum_{j\in \Omega_i}
R^{\alpha}R^{\beta} W(|\bol{r}_i-\bol{r}_j|,h).\label{eq:msph4}
\end{align}
All values in Eq. \eqref{eq:MSPH}, except for $\partial_{\alpha} f$
and $\partial_{\beta} \partial_{\gamma} f$,
are calculated using Eq. \eqref{eq:sph}, 
and Eqs. \eqref{eq:msph1}--\eqref{eq:msph4}.
Solving Eq. \eqref{eq:MSPH} using the usual linear algebra, 
we obtain $\partial_{\alpha} f$ ($\alpha=\{x,y,z\}$).
The square matrix in the right-hand side of Eq. \eqref{eq:MSPH} 
is common among the macroscopic variables 
in the same configuration $\{\bol{r}_i\}$.
Therefore, 
once the square matrix is decomposed using LU decomposition,
we can reuse the decomposed matrix,
decreasing the numerical cost 
of solving the linear equations (Eq. \eqref{eq:MSPH})
for variables in the same fluid particle configuration.

In Eqs. \eqref{eq:velocity} and \eqref{eq:position}, 
the value of a physical field 
at the position of a fluid particle is given by 
the physical quantity that the fluid
particle holds, and the derivatives of the physical field 
can be obtained through Eq. \eqref{eq:MSPH}. 
The density field, however, 
is not a physical variable that a fluid particle possesses 
but a physical variable determined from a configuration of fluid particles.
Thus, we use
the following definition for the density field
in the same manner as SPH,\cite{SPH}
\begin{equation}
\rho_i\equiv \sum_{j\in \Omega_i} m_j W(|\bol{r}_i-\bol{r}_j|,h),
\end{equation}
where $m_i$ is the mass of the $i$-th fluid particle.
Here, we impose incompressibility to the polymer melt.
To maintain the initial uniform density at each fluid particle,
we assume the pressure $p_i$ to be proportional to 
the local density $\rho_i$,\cite{MFZ1997}
\begin{equation}
p_i = 
c^2\rho_i,\label{eq:press}
\end{equation}
where $c$ is the sound velocity.
When the fluid particles fill all the space,
repulsive forces act between particles
through Eqs. \eqref{eq:velocity} and \eqref{eq:press}.
When we set $c=1.0 [a/t_0]$, 
where $t_0$ is the time unit in the Lagrangian fluid dynamics simulation,
the fluid is almost incompressible 
when $Re<1.0$,
and thus we use this value.

As mentioned above,
the stress tensor $\bol{\sigma}$ of the entangled polymer melt
is described by the dynamic response 
of the microscopic state of the polymers.
To efficiently obtain the stress tensor $\bol{\sigma}$ of
the entangled polymer melt,
we employ a coarse-grained model for the polymers,
i.e., the slip-link model\cite{SLTM2001,TTD2001,DT2003} that
can accurately describe 
the dynamics of entangled polymers. 
In a well-entangled polymer melt,
a polymer chain can be considered to be in a tube, 
composed of surrounding polymer chains.\cite{DE1986} 
In the slip-link model,
the slip-links substitute for the tube that constrains
the polymer from moving
perpendicular to the polymer contour direction
but allows the polymer to move parallel to the contour direction,
i.e., the reptation motion.\cite{DE1986}
Instead of the tube description,
representative constraints, i.e., slip-links, are traced 
in the slip-link model.

Focusing on a section of the polymer chain 
between two adjacent slip-links along a polymer,
both ends of the section are fixed by the slip-links,
and the chain conformation 
of the section
has a complex curve. 
Connecting the slip-links creates
a primitive path with a complex curve.
Therefore, the orientation of the part of the polymer chain
between two adjacent slip-links
is considered to be composed of the following: 
i) 
the orientation of the primitive path
that directly connects the adjacent slip-links 
(i.e., the average orientation of the section of the polymer chain
in the scale of the primitive path length)
and ii) a random segment orientation on a much shorter length scale
than that of the primitive path.
This random segment orientation can be regarded as isotropic 
because 
the dynamics of segments smaller
than the primitive path 
are much faster 
than the dynamics of the primitive path.
The stress tensor $\bol{\sigma}$ depends on the conformations of
the microscopic polymers,
and 
the stress tensor $\bol{\sigma}$ can therefore be  divided into
the following:
i) $\bol{\sigma}^{\rm p}$ from the primitive paths
and ii) $\bol{\sigma}^{\rm d}$ from the isotropic part.
The latter stress tensor is assumed to be $\bol{\sigma}^{\rm
d}=\eta_{\rm d}\bol{D}$, where $\bol{D}=\bol{\nabla v}+(\bol{\nabla v
})^{\rm T}$ because of its isotropy.

At each time step of the slip-link simulation,
the conformations of the slip-links are updated 
by the following four steps:
\begin{enumerate}
\item Affine deformation according to the macroscopic 
      local velocity gradient tensor $\bol{\kappa}\equiv
      (\bol{\nabla v})^{\rm T}$;
\item Change in the contour length of the primitive path
      due to the random motion of the slip-links;
\item Reptation motion along the contour 
      due to the random motion of the centre of mass of the polymer;
\item Constraint renewal because a slip-link is removed when the
      slip-link has slipped off from the polymer chain
      or created when the one of the ends of the polymer chain 
      entangles with the other polymer chain.
\end{enumerate}
For more details, see Refs. \onlinecite{TTD2001,DT2003}.
For a given chain configuration, the stress tensor 
is calculated by
\begin{equation}
\sigma_{\alpha \beta}^{\rm p} =
\sigma_{\rm e} 
\sum_{i=1}^{Z}
\left\langle
\frac{
r_{i\alpha}^{\rm s}r_{i\beta}^{\rm s}
}
{a_{\rm s}|\bol{r}_i^{\rm s}|}
\right\rangle,\label{eq:stress}
\end{equation}
where 
$a_{\rm s}$ is the unit length in the slip-link model and
$r_{i\alpha}^{\rm s}$ is the
$\alpha$-component of the $i$-th segment vector connecting
adjacent slip-links along a polymer. 
The number of slip-links on a chain is represented by $Z$.
The unit of stress $\sigma_{\rm e}$ in the slip-link model is
related 
to the plateau modulus $G_{\rm N}$ 
by $\sigma_{\rm e}=(15/4)G_{\rm N}$.\cite{DT2003}
The slip-link model has two characteristic timescales: 
the Rouse relaxation time
$\tau_{\rm R}$ and the longest relaxation time $\tau_{\rm d}$.
These characteristic times are
related to $Z$ as follows:
$\tau_{\rm R}=Z^2 t_{\rm e}$ 
and $\tau_{\rm d}\propto Z^{3.4}t_{\rm e}$,
where $t_{\rm e}$ is the time unit in the slip-link simulation.\cite{DT2003} 
The contour-length relaxation of a confining tube,
i.e., the extensional relaxation,
occurs on the timescale of $\tau_{\rm R}$,
while the orientation relaxation occurs on the timescale of $\tau_{\rm d}$.

Each polymer simulator describing a fluid particle 
computes the polymer configurations
at each time step, 
and the recorded configurations are used as the initial states in
the next time step.
Typically, the macroscopic time unit $t_{\rm macro}$
and microscopic time unit $t_{\rm micro}$
have a large timescale gap, and 
the macroscopic time unit $t_{\rm macro}$ must therefore be divided
into $N t_{\rm micro}$.
Because the slip-link model used here is sufficiently coarse-grained 
and the time unit $t_{\rm e}$ 
can be the same as the timescale of the macroscopic fluid
$t_{\rm macro}=t_0$,
we employ 
$t_{\rm macro}=t_{\rm micro}\equiv t_{\rm e}$.

\subsection{Two-dimensional polymeric flow with memory advection}
To demonstrate the efficiency of the presented multiscale simulation,
we consider a two-dimensional polymer melt system 
in which the flow history can affect the flow behaviour.
Figure \ref{fig:conv_flow} shows the system 
in which a circular object with radius $r_{\rm c}=3a$ is fixed on
the centre of the square system with sides $2H$, 
where $H$ is set to be $H=15a$.
We imposed the stick boundary condition on the velocity
at the perimeter of the circular object
and periodic boundary conditions 
on all the physical variables used 
at the boundaries of the system.
Around the circular object, the polymers in a fluid particle
experience a strain rate that depends on the position
along the stream line,
and the conformations of the polymers
turn out to be
different between the upstream and downstream regions.
The difference in the conformations between these two regions
can be observed macroscopically in the distribution of the stress 
through Eq. \eqref{eq:stress}.

\begin{figure}
\begin{center}
\includegraphics[scale=0.5]{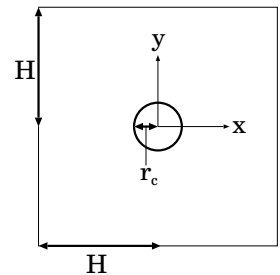}
\end{center}
\caption{\label{fig:conv_flow}
Two dimensional polymeric flow system with
a circular obstacle. 
The circular object with a radius $r_{\rm c}$ is fixed on the centre of 
a square system with sides $2H$.
To describe the circular object,
fixed particles are evenly placed
on the circle.
Mobile fluid particles flow outside of the circle.
The inside of the circle is vacant.
Periodic boundary conditions are imposed at the sides of the square system.
}
\end{figure}

The polymer melt is composed of monodisperse linear polymers
with an average number of entanglements 
on each polymer of $Z=7$.
In the bulk of this system, 
the zero-shear viscosity $\eta_0^{\rm p}$ and the
longest relaxation time $\tau_{\rm d}$ are estimated 
from the slip-link simulation to be 
$\eta_0^{\rm p}\simeq 17.5 \sigma_{\rm e}t_{\rm e}$ and $\tau_{\rm
d}=200 t_{\rm e}$.\cite{MT2010b}
Approximately 900 fluid particles are evenly placed in the system.
The circular object fixed in the space 
is expressed by 24 fluid particles
placed at 
the perimeter of the circle; 
the inside of the circle is vacant.
Each fluid particle consists of 1000 polymers,
enough to describe the bulk rheological properties of 
a single fluid particle.\cite{MT2010a}
Under the body force 
$\bol{F}^{\rm b}/(\eta^0/at_{\rm e}) = (5.0\times 10^{-4}, 0)$,
the flow becomes a steady-state in about 500$t_{\rm e}$.
The average steady-state flow velocity in this system 
is nearly equal to $(0.055, 0) [a/t_{\rm e}]$ 
for a polymer melt flow with $Z=7$ and $\eta_{\rm d}/\eta^0=0.1$
and the Reynolds number $Re = \rho U r_{\rm c}/\eta^0$ is 0.2.
Under these conditions, the flow is almost laminar, i.e., 
the flow between the upstream and downstream regions
can be symmetric at the steady-state 
of low Reynolds number flow
in
the Newtonian fluid.
In the regions where $D_{xy}> 1/\tau_{\rm d}$,
the Weissenberg number $We$ is larger than 1,
and the flow shows shear-thinning.
The shear-thinning behaviour is caused by 
the anisotropic distribution of the polymer chain orientation,
i.e., the memory of the strain rate
that the polymer chains have experienced.
The orientation relaxation is not accomplished
in a flow with a strain rate larger than $1/\tau_{\rm d}$ 
within a time interval less than $\tau_{\rm d}$.
Similar to the orientation relaxation,
the extensional relaxation is not accomplished
in a flow with a strain-rate flow 
larger than $1/\tau_{\rm R}$ 
within a time less than $\tau_{\rm R}$.
Because the Rouse relaxation time $\tau_{\rm R}$
is much smaller than the longest relaxation
time $\tau_{\rm d}$, 
the extensional relaxation progresses easily compared to 
the orientation relaxation, which is obstructed by entanglements.
Therefore,
the extensional relaxation is difficult to observe
in the polymer melt flow without careful treatment and observation.
In the present multiscale simulation 
following the Lagrangian fluid dynamics at the macroscopic level,
we can consider 
not only the orientation relaxation
but also the extensional relaxation
caused by the memory of the strain rate
because the microscopic simulators are composed of entangled polymers
and the advection of the entire polymer configurations
is involved.

Figures \ref{fig:conv_tra} and \ref{fig:conv_tra_N}
show the transient flow behaviour of magnitudes of 
(a) velocity $|\bol{v}| [a/t_{\rm e}]$,
(b) strain rate $|D_{xy}| [1/t_{\rm e}]$,
and (c) shear stress over zero-shear viscosity 
$|\sigma_{xy}|/\eta^0 [1/t_{\rm e}]$
at $t = 100, 200, 400$,
and $800 [t_{\rm e}]$
in the polymer melt and the Newtonian fluid, respectively.
To obtain these contour maps,
we performed a linear interpolation to transform 
the data at the particle positions into values 
at regular lattice points.
To decrease the spiky noise of the data, 
median-filtering,
replacing the value with the median value
obtained from the values of the neighbouring pixels 
and the pixel's own value,
was carried out to
the evaluated data on the regular lattice points.
Comparing Figs. \ref{fig:conv_tra} (c) and
Figs. \ref{fig:conv_tra_N} (c),
an asymmetry in the polymeric stress 
between the upstream and downstream regions is apparent.
As seen in Fig. \ref{fig:conv_tra} (c),
the asymmetry in the polymeric stress 
between the upstream and downstream regions
develops gradually, while such asymmetry developments
do not appear in the velocity and strain-rate field,
as shown in Figs. \ref{fig:conv_tra} (a) and (b).
The Weissenberg number $We$ is larger than unity in the vicinity
of the circle as shown in Fig. \ref{fig:conv_tra} (b) by thick lines.
A nonlinear relationship is observed between $D_{xy}$ and $\sigma_{xy}$
in the regions where $We > 1$ 
and even when $We \le 1$.
\begin{figure*}
\begin{center}
\includegraphics[scale=0.8]{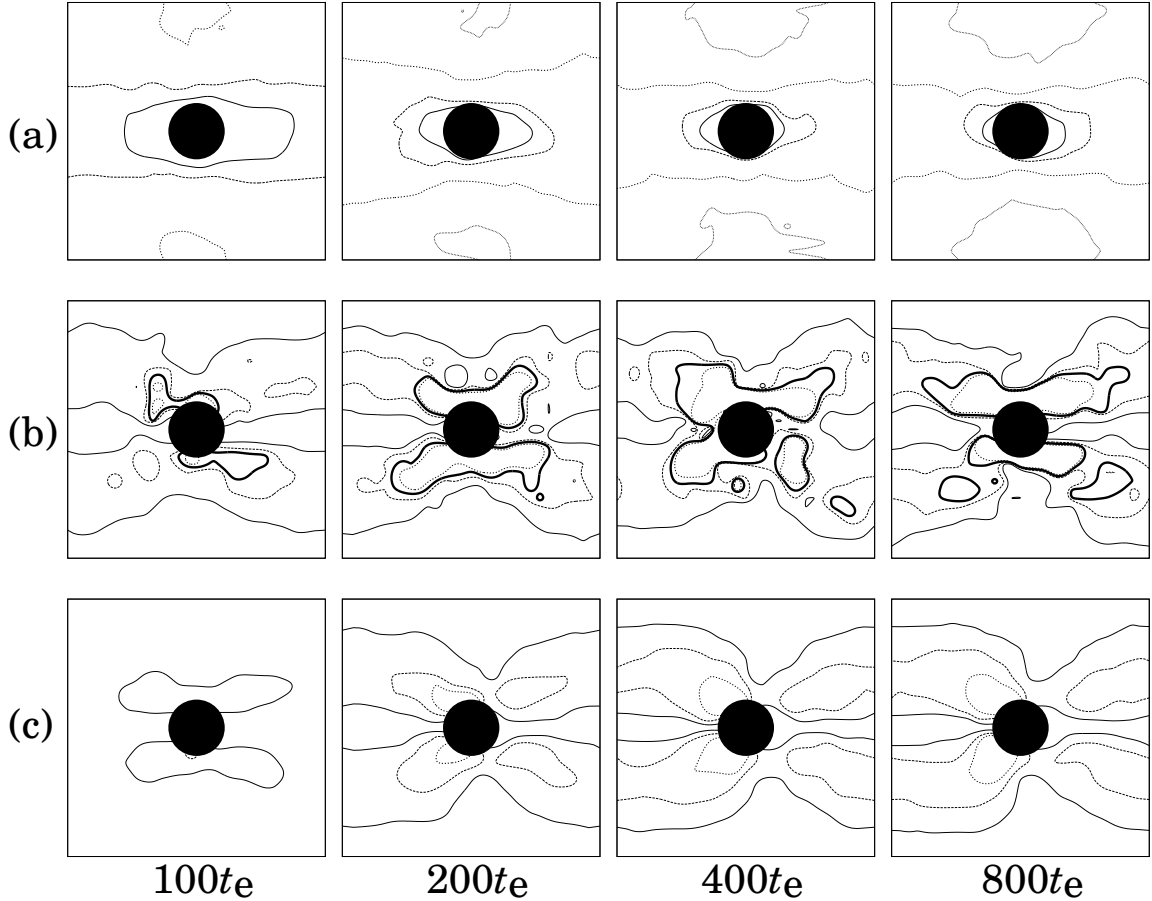}
\end{center}
\caption{\label{fig:conv_tra}
Transient flow behaviour in the viscoelastic polymer melt.
Contour maps of magnitudes of (a) velocity $|\bol{v}| [a/t_{\rm e}]$, 
(b) strain rate $|D_{xy}| [1/t_{\rm e}]$,
and (c) shear stress over zero-shear viscosity 
$|\sigma_{xy}|/\eta^0 [1/t_{\rm e}]$ 
at 100, 200, 400, and 800 $[t_{\rm e}]$ are presented.
Contour levels of these figures 
are 0.02 (solid line), 0.03 (broken line),
0.05 (dotted line), and 0.07 (dash-dot line) $[a/t_{\rm e}]$ in (a),
and 0.002 (solid line), 0.004 (broken line),
0.006 (dotted line) $[1/t_{\rm e}]$ in (b) and (c).
The thick solid lines in (b) indicate $We=1$.
}
\end{figure*}

\begin{figure*}
\begin{center}
\includegraphics[scale=0.8]{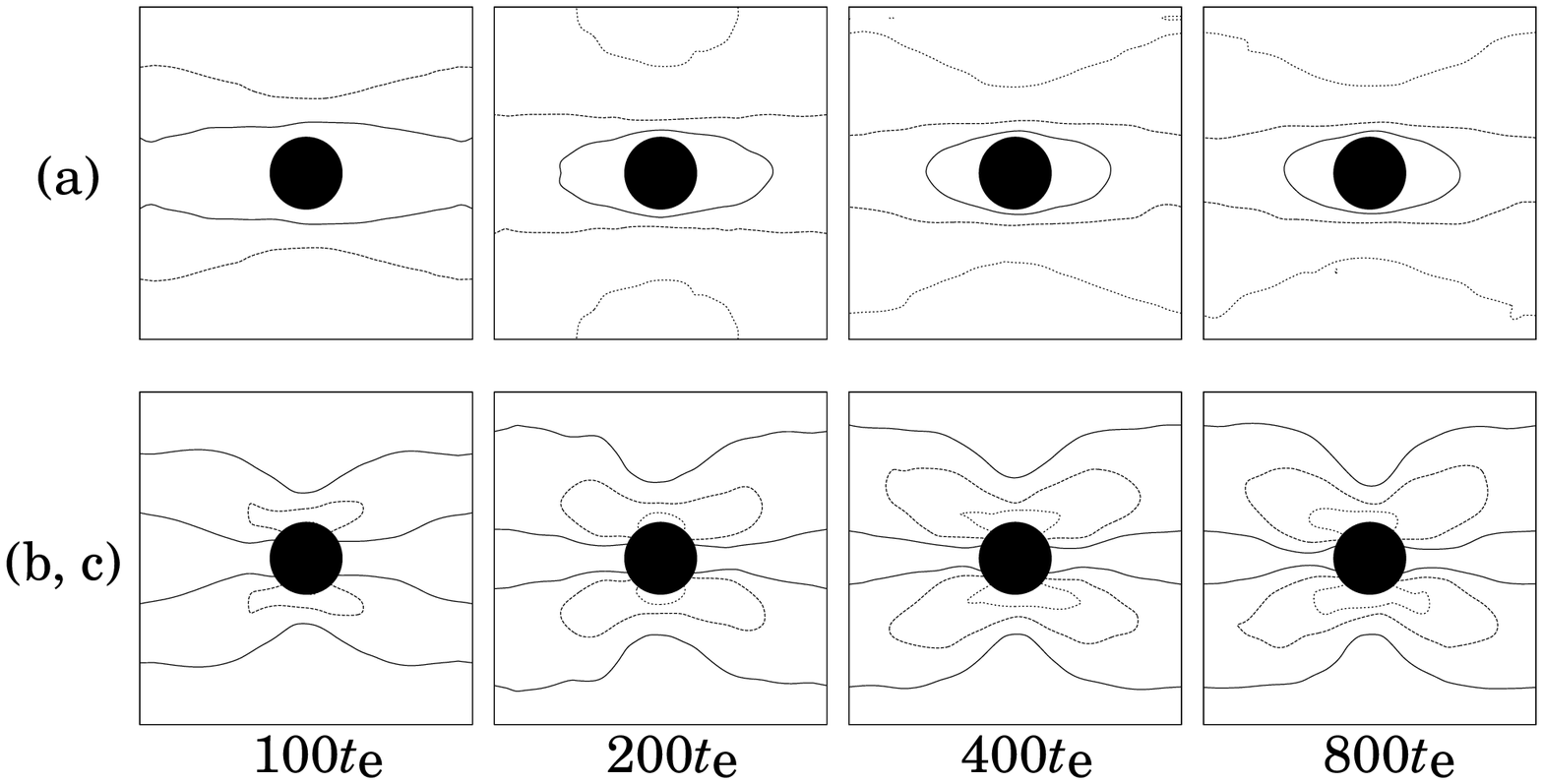}
\end{center}
\caption{\label{fig:conv_tra_N}
Transient flow behaviour in the Newtonian fluid.
Contour maps of magnitudes of (a) velocity $|\bol{v}| [a/t_{\rm e}]$, 
(b) strain rate $|D_{xy}| [1/t_{\rm e}]$,
and (c) shear stress over zero-shear viscosity 
$|\sigma_{xy}|/\eta^0 [1/t_{\rm e}]$ 
at 100, 200, 400, and 800 $[t_{\rm e}]$ are presented.
In case of the Newtonian fluid,
Figures (b) and (c) are abbreviated to the same figures
because $|D_{xy}|=|\sigma_{xy}|/\eta^0$. 
Contour levels of these figures 
are 0.02 (solid line), 0.03 (broken line),
0.05 (dotted line) $[a/t_{\rm e}]$ in (a),
and 0.002 (solid line), 0.004 (broken line),
0.006 (dotted line) $[1/t_{\rm e}]$ in (b) and (c).}
\end{figure*}

To highlight
the nonlinear relationship between the strain rate $D_{xy}$
and the shear stress $\sigma_{xy}$
in the asymmetric stress field shown in Fig. \ref{fig:conv_tra} (c),
we consider the deviation of the polymeric shear stress $\sigma_{xy}$
from the Newtonian stress estimated using 
the zero-shear-rate viscosity of the polymer melt 
$\sigma_{xy}^0\equiv\eta^0 D_{xy}$, as follows:
\begin{equation}
\Delta \sigma_{xy}\equiv |\sigma_{xy}| - |\sigma_{xy}^0|.
\end{equation}
Figure \ref{fig:stress_overshoot} shows the cases of
(a) $\Delta \sigma_{xy} \ge 0$ and (b) $\Delta \sigma_{xy} < 0$
at $t=800 [t_{\rm e}]$.
When the stress deviation $\Delta \sigma_{xy}(\bol{r})$ is 
positive at a certain position $\bol{r}$,
the stress exhibits an overshoot coming from 
the elastic contribution of unoriented but well-extended polymer
chains.
When $\Delta \sigma_{xy}(\bol{r})$ is 
negative, on the other hand,
the stress represents a shear-thinning behaviour. 
As seen in Fig. \ref{fig:stress_overshoot} (a),
stress overshoots are clearly observed in the upstream regions;
these are
indicated by the white dashed circles in the figure.
In the regions where $We > 1$,
the orientations of polymer chains cannot recover the isotropy.
In the regions where $We > 1$,
the inside of the solid lines in Fig. \ref{fig:stress_overshoot} (b),
we can clearly see the shear-thinning behaviour arising from the nonlinear
relationship between the shear-rate and shear stress, e.g.,
$\sigma_{xy}\propto D_{xy}^n$ $(0<n<1)$.
Shear-thinning behaviour is also observed in the regions where
$We \le 1$, indicated by the black dashed circles in
Fig. \ref{fig:stress_overshoot} (b).
In the regions where $We \le 1$,
the orientations of polymer chains 
are expected to be isotropic,
and the stress tensor should thus be proportional to the strain-rate tensor.
However, the stress tensor is not proportional to
the strain-rate tensor even in the regions where $We \le 1$.
Instead,
nonlinear behaviours, i.e.,
stress overshoot and shear-thinning behaviour,
appear in the regions where $We \le 1$
because of the memory effect 
relating to the extensional relaxation.
\begin{figure*}
\begin{center}
\includegraphics[scale=1]{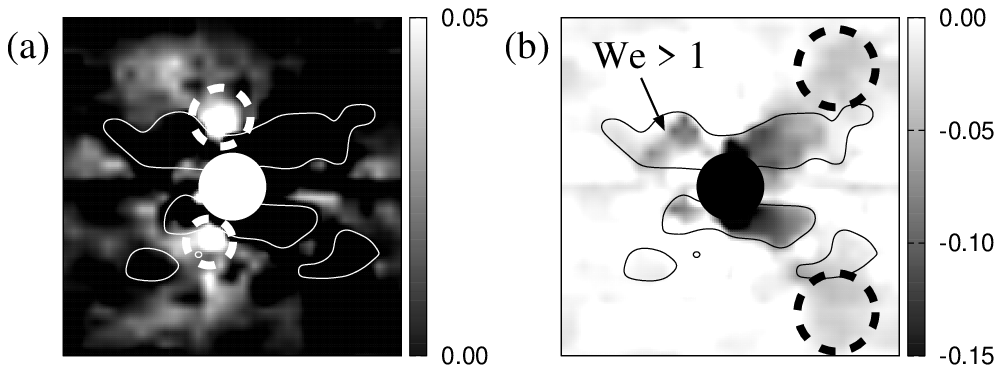}
\end{center}
\caption{\label{fig:stress_overshoot}
Greyscale contour maps of
$\Delta \sigma_{xy}\equiv
|\sigma_{xy}|-|\sigma_{xy}^0| [\sigma_{\rm e}]$ 
at $t=800 [t_{\rm e}]$
are separately presented
for cases in which
the values of $\Delta \sigma_{xy}$ are (a) positive and (b) negative.
The regions surrounded by 
the solid lines (white in (a) and black in (b))
represent the regions in which $We > 1$.
The dashed circles (white in (a) and black in (b))
indicate regions where the stress overshoot and shear-thinning behaviour 
caused by the memory effect in the stress
are clearly observed in (a) and (b), respectively.}
\end{figure*}

To quantitatively clarify the asymmetry of $|\bol{v}|$,
$|D_{xy}|$ and $|\sigma_{xy}|$
with respect to the $y$-axis shown in Fig. \ref{fig:conv_flow},
we define a measure of asymmetry in a field $f$
as
\begin{align}
{\rm Asym}(f)\equiv
\frac{\sum_{x=-H/a}^{H/a}\sum_{y=-H/a}^{H/a} 
||f(x,y)|-|f({\bar{x}},y)||}
{(H/a+1)(H/a+1) f_{\rm MAX}^{\infty}},
\end{align}
where 
$x$ and $y$ are the lattice coordinates in $-H/a \le x \le H/a$ 
and $-H/a \le y \le H/a$, respectively,
$\bar{x}\equiv-x$.
The maximum value of ${\rm Asym}(f)$ at the steady-state 
$f_{\rm MAX}^{\infty}$
is defined as 
$f_{\rm MAX}^{\infty}=\int_{t>t_{\rm s}} |f_{\rm MAX}(t)| 
dt/\int_{t>t_{\rm s}}dt$ where $t_{\rm s}$ is the time when
the system reaches the steady-state and
$f_{\rm MAX}(t)$ is the maximum value of $f$ in the system 
at the time $t$. 

Figure \ref{fig:asym} shows the time evolutions of 
(a) ${\rm Asym}(|\bol{v}|)$, (b) ${\rm Asym}(|D_{xy}|)$,
and (c) ${\rm Asym}(|\sigma_{xy}|)$ represented with dotted line,
broken line, and solid line, respectively,
for the polymer melt (P) with $Z=7$ and $\eta_{\rm d}/\eta^0=0.1$
and the Newtonian fluid (N) with $\eta^0=\eta_{\rm d}$.
Transient behaviour can be seen in Fig. \ref{fig:asym} before 
$t=400 [t_{\rm e}]$ in the polymer melt
and $t=200 [t_{\rm e}]$ in the Newtonian fluid.
After $t=400 [t_{\rm e}]$ in the polymer-melt case
(or $t=200 [t_{\rm e}]$ in the Newtonian case), these asymmetry measures 
randomly oscillate around each mean value in the steady-state.
Thus, the time in steady-state $t_{\rm s}$ is estimated 
to be $400 [t_{\rm e}]$ in the polymer-melt case
and $200 [t_{\rm e}]$ in the Newtonian case. 
The asymmetry measure of $|\sigma_{xy}|$ of the polymer melt
shows a typical time-evolution behaviour, i.e.,
the flow history develops an asymmetry in the stress field. 
As shown in Fig. \ref{fig:asym},
the asymmetry measure of $|\sigma_{xy}|$ of the polymer melt
is much higher than that of
$|\bol{v}|$ and $|D_{xy}|$ in the steady-state.
Reflecting the larger noise in Fig. \ref{fig:conv_tra} (b)
than in Fig. \ref{fig:conv_tra} (a),
the asymmetry measure of $|D_{xy}|$ of the polymer melt
is slightly higher than that of
$|\bol{v}|$ in the steady-state. 
The noise of the data raises the asymmetry measure.
Even after accounting for the noise of the data,
the asymmetry measure of $|\sigma_{xy}|$ of the polymer melt
is much higher than that of the others.
The asymmetry measures of the Newtonian fluid at the steady-state
have small values as shown in Fig. \ref{fig:asym} (N). 
This small asymmetry is caused by the asymmetric distribution 
of the positions of fluid particles
constituting the Lagrangian fluid.
The asymmetry measures of $|\bol{v}|$ and $|D_{xy}|$
of the polymer melt are slightly higher than 
those of the Newtonian fluid.
The velocity and strain rate do not directly reflect
memory of the flow history, although the memory indirectly affects
them through the macroscopic stress, which
is influenced by the memory of the strain rate.
Therefore, the asymmetry measures of both $|\bol{v}|$ and $|D_{xy}|$ are
weaker than that of $|\sigma_{xy}|$ in the polymer-melt case.
\begin{figure*}
\begin{center}
\includegraphics[scale=1]{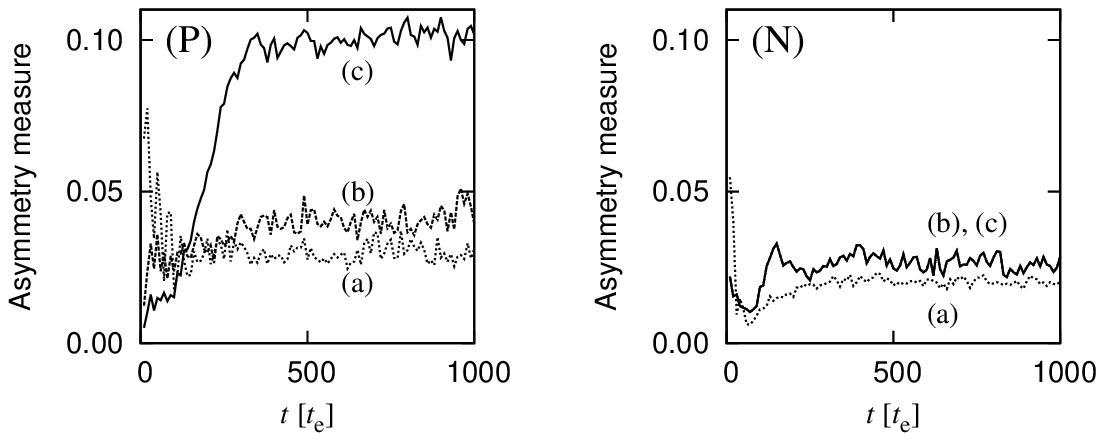}
\end{center}
\caption{\label{fig:asym}Time evolution of asymmetry measure 
of (a) $|\bol{v}|$ (dotted line), 
(b) $|D_{xy}|$ (broken line), and
(c) $|\sigma_{xy}|$ (solid line)
for (P) polymer melt flow with $Z=7$ and $\eta_{\rm d}/\eta^0=0.1$,
and (N) Newtonian flow with $\eta^0=\eta_{\rm d}$.
}
\end{figure*}

The asymmetry in the stress distribution is caused by 
the strain-rate-history-dependent stress, 
i.e., the memory effect.\cite{MT2010b}
In Ref. \onlinecite{MT2010b}, we discussed the relationship
between the stress relaxation time of a polymer melt
in the homogeneous bulk and the retardation time of the stress field
from the strain-rate field.
We found that
the retardation time is comparable to the stress relaxation time
that actually corresponds to the Rouse relaxation time,
i.e., the extensional relaxation time in the entangled polymer melt.

\section{summary and future perspectives}
We have developed multiscale bridging simulations for simple liquids, one-dimensional polymeric liquids, and two-dimensional polymeric liquids. 
In our multiscale modelling, the macroscopic flow behaviours are calculated using the CFD method. However, instead of using any constitutive equations, a local stress is generated using a microscopic (or mesoscopic) simulation, e.g., MD simulation and coarse-grained polymer dynamics, associated with a local fluid element according to local flow variables obtained by the CFD calculation. 
This type of multiscale modelling is expected to be useful in predicting the macroscopic behaviours of complex fluids such as polymeric liquids and other complex softmatters in which the constitutive relations are unknown.

A multiscale simulation method of MD and CFD for simple fluids is developed in Section II. 
In this method, a usual finite-volume scheme with a lattice mesh is used for the CFD level, and each lattice is associated with a small MD cell that generates a ``local stress'' according to a ``local flow field'' obtained from the CFD calculation. 
The bridging scheme of CFD and MD is based on the local equilibrium assumption without memory (i.e., a local stress immediately attains the steady state after a short transient time according to a given flow field) [Fig. \ref{sec2_f00}]. 
The cavity flows of a Lennard-Jones liquid in a square box are calculated for large efficiency factors, e.g., $\Delta x/l_{\rm MD}$=$\Delta t/\tau_{\rm MD}$=8, and the results are compared with those of the Newtonian fluid and fluctuating hydrodynamics. 
With these efficiency factors, multiscale simulations are performed $8^{2}\times 8$ times faster than the full MD simulation. 
It is demonstrated that, although the instantaneous profiles of macroscopic quantities involve large fluctuations for large efficiency factors, the fluctuations are perfectly smoothed out by taking the time averages [Figs. \ref{sec2_1d_y_compari} and \ref{sec2_1d_x_compari}]. 
The spectrum analyses of fluctuations are carried out, and the fluctuations arising in the multiscale simulations are found to be consistent with those of the fluctuating hydrodynamics [Figs. \ref{sec2_f6} and \ref{sec2_f7}]. 
We also found a relation between the fluctuation and efficiency factor as Eq. (\ref{sec2_eq15}).

In Sec. III, the multiscale simulation method for simple fluids is extended to  one-dimensional flows of polymeric liquids in parallel plates, in which the macroscopic quantities are assumed to be uniform in the flow direction parallel to the plates. 
This assumption allows us to not trace the advection of memory of polymer configurations in each fluid element but instead to calculate the coupling of macroscopic flow behaviours to the microscopic dynamics of polymer chains using the usual fixed-mesh system at the CFD level. 
As in the simple fluids, a local stress is calculated by the non-equilibrium MD simulation in a small MD cell associated with a mesh interval of the CFD calculation. 
However, the MD simulations are performed in the time-step duration of the CFD calculation at each time step of CFD, and the final configuration of polymers obtained in each MD cell is retained as the initial configuration in each MD cell at the next time step. 
Thus, the memory of configurations of polymer chains is traced at the MD level so that the memory effects are correctly reproduced [Fig. \ref{sec3_f1}]. 
The behaviour of a glassy polymer melt in rapidly oscillating plates is calculated for various amplitudes and oscillation frequencies, and the velocity profiles and local rheological properties of the melt are investigated. 
The results of the multiscale simulations for a plate with an oscillation of small amplitude are also compared with those of the analytical solution for a plate with an infinitesimally small amplitude oscillation so as to demonstrate the validity of our multiscale simulations. 
The velocity profiles become thinner as the amplitude of oscillation of the plates increases due to the shear thinning occurring near the oscillating plate, where the local strain rates are large [Figs. \ref{sec3_f3} and \ref{sec3_f4}]. 
It is also demonstrated that the velocity profiles of the melt obtained by the multiscale simulations approach that of the linear analytical solution as the amplitude of oscillation of the plates decreases [Figs. \ref{sec3_f5} and \ref{sec3_f6}]. 
The local rheological properties are investigated in terms of the complex viscosity [Fig. \ref{sec3_f7}]. 
The local complex viscosities are demonstrated to become uniform and to approach those of the linear response as the amplitude of oscillation decreases. 
However, for large amplitudes, the local rheological properties are quite spatially varied. 
It is found that at high oscillation frequency and large amplitudes of oscillation of the plates, the melt has three different rheological regimes, i.e., viscous liquid, viscoelastic liquid, and viscoelastic solid regimes.

In Sec. IV, the multiscale simulation method, composed of a Lagrangian fluid dynamics simulation and a coarse-grained polymer simulation, is presented and applied to the entangled polymer-melt flow, in which the advection of microscopic variables affects its flow behaviour.
The Lagrangian fluid dynamics simulation enables us to treat the strain-rate-history-dependent stress field that is evaluated by a large number of microscopic simulations.
A local stress of the entangled polymer melt is obtained by a slip-link model that can efficiently calculate the dynamic response of the entangled polymer melt to an imposed strain rate in the homogeneous bulk.
As a demonstration, we consider the transient flow of the polymer melt passing a circular object in a two-dimensional periodic system.
In order to use the Lagrangian fluid dynamics simulation at the macroscopic level, we can trace the polymer conformations in the inhomogeneous flow field, and show that the effect of memory that is preserved as the polymer conformations in a fluid particle appears as a macroscopic stress distribution with an asymmetry between the upstream and downstream regions around the circular object [Fig. \ref{fig:conv_tra}].
Nonlinear behaviours, represented by stress overshoot and shear-thinning behaviours, are observed in the regions where $We \le 1$ because of the memory effect relating to the extensional relaxation [Fig. \ref{fig:stress_overshoot}].
To quantify the asymmetry between the upstream and downstream regions around the circular object, we define an asymmetry measure and investigate its transient behaviour.
The asymmetry measure of the polymeric stress is found to show typical transient behaviours: i) the asymmetry measure of the stress increases before $t=t_{\rm s}$ and ii) the asymmetry measure approaches a steady-state value after $t=t_{\rm s}$, while the asymmetry measure of the velocity and the strain rate do not represent transient behaviour [Fig. \ref{fig:asym}].

In the present work, we have developed a multiscale simulation of the multi-dimensional flows of polymeric liquids involving the advection of memory. 
Tracing of the memory and its advection is quite important not only in polymeric flows but also in many kinds of softmatter flows. 
In this work, we discuss only isothermal fluids with a uniform density. 
We have also assumed the non-slip boundary condition for the macroscopic velocity at the wall. 
Thus, there remain two important directions of development of multiscale modelling for the complex flows of softmatters. 
One is the treatment of the couplings of various macroscopic quantities, e.g., temperature, density, and velocity. In particular, the inclusion of the temperature variation may be important for polymeric liquids. 
The other important developmental direction involves consideration of the role of microscopic dynamics at the interface in determining macroscopic flow behaviours. 
Some simulation methods have been proposed for this purpose.\cite{art:95OT,art:00FWF,art:03DC,art:05DFC}
Incorporating one of these into our multiscale simulation method may enable us to treat slip motion, adhesion, and anchoring at the interface together with the macroscopic flow behaviours. 
Furthermore, although the high computational expense of describing three-dimensional systems prevents us from simulating the three-dimensional flows, the presented multiscale formulations allow consideration of three-dimensional flows of soft materials in principle. 
In the near future, developments in computer architecture and more parallelisation should enable us to account for such dense multiscale systems.



\begin{thebibliography}{100}
\bibitem{note1}
In the present simulations of fluctuating hydrodynamics, the random noises are included only in the off-diagonal components of the stress tensor $T'$. That is, we put $T'_{xx}=T'_{yy}=0$ and 
$T'_{xy}$=$T'_{yx}$=$2\eta E'_{xy}$+$R'_{{\rm CFD}{xy}}$,
where the intensity of the random noise is defined by the Fluctuation-dissipation theorem as $\langle R_{{\rm CFD}{xy}}'^2\rangle = 2\eta kT/(\Delta x^3\Delta t)$.
\end{thebibliography}
%
%

\end{document}